\documentclass{aa}
\usepackage[utf8]{inputenc}
\usepackage[varg]{txfonts}
\usepackage[breaklinks, colorlinks, citecolor=blue, linkcolor=blue]{hyperref}
\usepackage[usenames, dvipsnames]{color}
\usepackage{amsmath}
\usepackage{graphicx}
\usepackage[english]{babel}
\usepackage{siunitx}
\usepackage{float}
\usepackage{url}
\usepackage{longtable}
\usepackage{fancyhdr}
\usepackage{footmisc}
\usepackage{natbib}
\usepackage[normalem]{ulem}
\usepackage{caption}
\usepackage{multirow}
\usepackage{color}
\usepackage{array} 
\usepackage{ulem}
\usepackage[flushleft]{threeparttable}
\usepackage{orcidlink}
\makeatletter
\def\instrefs#1{{\def\scsep{\def\scsep{,}}\@for\w:=#1\do{\scsep\ref{inst:\w}}}}
\renewcommand{\inst}[1]{\unskip$^{\instrefs{#1}}$}

\renewcommand*\aa@pageof{, page \thepage{} of \pageref*{LastPage}} % https://tex.stackexchange.com/questions/345764/journal-class-shows-package-hyperref-warning-suppressing-link-with-empty-targe

\makeatother
\begin{document}
%-------------------------- TITLE --------------------------%

%Boring planets also deserve to be observed II:
\title{The obliquity and atmosphere of the hot Jupiter WASP-122b (KELT-14b) with ESPRESSO: An aligned orbit and no sign of atomic or molecular absorption\thanks{Based on Guaranteed Time Observations collected at the European Southern Observatory under ESO program 106.21M2.004 by the ESPRESSO Consortium.}}

\titlerunning{Obliquity and atmosphere of the hot Jupiter WASP-122b (KELT-14b) with ESPRESSO}

\author{{M.~Stangret\inst{inaf_padova}\orcidlink{0000-0002-1812-8024}}
        \and 
        E.~Palle
        \inst{iac,ull}
        \orcidlink{0000-0003-0987-1593} %checked the paper
        \and
        E.~Esparza-Borges \inst{iac,ull}\orcidlink{0000-0002-2341-3233}
        \and
        J.~Orell Miquel \inst{iac,ull}\orcidlink{0000-0003-2066-8959} %comments received
        \and
        N.~Casasayas-Barris\inst{iac}\orcidlink{0000-0002-2891-8222}
         \and
        M.~R.~Zapatero Osorio\inst{madrid_csic}\orcidlink{0000-0001-5664-2852}
        \and    
        E.~Cristo\inst{porto_instit,porto_depart}\orcidlink{0000-0001-5992-7589} %comments send
        \and      
        R.~Allart\inst{montreal,obs_geneva}\orcidlink{0000-0002-1199-9759} %comments send, acknowledgements send I need to add them
        \and
        Y. Alibert\inst{bern}
        \and
        F.~Borsa\inst{INAF_Brera}\orcidlink{0000-0003-4830-0590} %comments send
%comments send
        \and   
        O.~D.~S.~Demangeon\inst{porto_instit,porto_depart}\orcidlink{0000-0001-7918-0355}
        \and
        P.~Di~Marcantonio\inst{inaf_trieste}\orcidlink{0000-0003-3168-2289} %no comments - co_A, Orcid send
        \and
        D.~Ehrenreich\inst{obs_geneva,center_geneva}\orcidlink{0000-0001-9704-5405} %comments send, affiliation send
        \and
        P.~Figueira\inst{obs_geneva,porto_instit} %comments send, Geneva 
        \and
        J. I. Gonz\'alez Hern\'andez\inst{iac,ull} %no comments - co-A
         \and
        E.~Herrero-Cisneros\inst{CSIC_Torrejon}\orcidlink{0000-0002-9990-6915}
        \and    
        C.~J.~A.~P.~Martins\inst{porto_centro,porto_instit}\orcidlink{0000-0002-4886-9261} %no comments - co_A+ acknowledgements
        \and
        N.~C.~Santos\inst{porto_instit,porto_depart}
        \and
        J.~V.~Seidel\inst{ESO_Chile}\orcidlink{0000-0002-7990-9596} %comments send
        \and
        T.~Azevedo~Silva\inst{porto_instit,porto_depart,INAF_Arcetri}\orcidlink{0000-0002-9379-4895}
        \and
        A.~Sozzetti\inst{INAF_Torino}\orcidlink{0000-0002-7504-365X} %comments send
        \and
        M. Steiner\inst{obs_geneva}
        \and
        A.~Suárez Mascareño\inst{iac,ull}\orcidlink{0000-0002-3814-5323}
        \and 
        S.~Udry\inst{obs_geneva}\orcidlink{0000-0001-7576-6236} %something send
        }

\institute{
\label{inst:inaf_padova}INAF – Osservatorio Astronomico di Padova, Vicolo dell'Osservatorio 5, 35122, Padova, Italy \\ \email{monika.beata.stangret@gmail.com}%1
\and
\label{inst:iac}Instituto de Astrof\'isica de Canarias (IAC), 38205 La Laguna, Tenerife, Spain %2
\and 
\label{inst:ull}Departamento de Astrof\'isica, Universidad de La Laguna (ULL), 38206, La Laguna, Tenerife, Spain %3
\and
\label{inst:madrid_csic}Centro de Astrobiología, CSIC-INTA, Camino Bajo del Castillo s/n, 28692, Villanueva de la Cañada, Madrid, Spain %4
\and
\label{inst:porto_instit}Instituto de Astrof\'{\i}sica e Ci\^encias do Espa\c co, CAUP, Rua das Estrelas, 4150-762 Porto, Portugal %5
\and
\label{inst:porto_depart}Departamento de F\'{\i}sica e Astronomia, Faculdade de Ci\^encias, Universidade do Porto, Rua Campo Alegre, 4169-007 Porto, Portugal %6
\and
\label{inst:montreal}D\'epartement de Physique, Institut Trottier de Recherche sur les Exoplan\`etes, Universit\'e de Montr\'eal, Montr\'eal, Qu\'ebec, H3T 1J4, Canada %7
\and
\label{inst:obs_geneva}Observatoire Astronomique de l'Universit\'e de Gen\`eve, Chemin Pegasi 51, Sauverny, CH-1290, Switzerland %8
\and
\label{inst:bern}Physics Institute of University of Bern, Gesellschafts strasse 6, 3012, Bern, Switzerland %9
\and
\label{inst:INAF_Brera}INAF – Osservatorio Astronomico di Brera, Via Bianchi 46, 23807 Merate, Italy %10
\and
\label{inst:inaf_trieste}INAF – Osservatorio Astronomico di Trieste, via G. B. Tiepolo 11, I-34143, Trieste, Italy %11
\and
\label{inst:center_geneva}Centre Vie dans l'Univers, Facult\'e des sciences de l'Universit\'e de Gen\`eve, Quai Ernest-Ansermet 30, 1205 Geneva, Switzerland %12
\and
\label{inst:porto_centro}Centro de Astrof\'{\i}sica da Universidade do Porto, Rua das Estrelas, 4150-762 Porto, Portugal %13
\and
\label{inst:ESO_Chile}European Southern Observatory (ESO), Alonso de Córdova 3107, Vitacura, Casilla 19001, Santiago de Chile, Chile %14
\and
\label{inst:INAF_Torino}INAF – Osservatorio Astrofisico di Torino, Via Osservatorio 20, 10025 Pino Torinese, Italy %15
\and
\label{inst:CSIC_Torrejon}Centro de Astrobiolog\'ia, CSIC-INTA, Crta. Ajalvir km 4, E-28850 Torrej\'on de Ardoz, Madrid, Spain 
\and
\label{inst:INAF_Arcetri}INAF - Osservatorio Astrofisico di Arcetri, Largo Enrico Fermi 5, I-50125 Firenze, Italy
}

\date{}

\abstract{Thanks to their short orbital periods and hot extended atmospheres, hot Jupiters are ideal candidates for  atmosphere studies with high-resolution spectroscopy. New stable spectrographs help improve our understanding of the evolution and composition of those types of planets. By analyzing two nights of observations using the ESPRESSO high-resolution spectrograph, we studied the architecture and atmosphere of hot Jupiter WASP-122b (KELT-14b). By analyzing the Rossiter-McLaughlin (RM) effect, we measured the spin-orbit angle of the system to be $\lambda = 0.09^{+0.88}_{-0.90}$~deg. This result is in line with literature obliquity measurements of planetary systems around stars with effective temperatures cooler than 6500~K. 
Using the transmission spectroscopy, we studied the atmosphere of the planet. Applying both the single-line analysis and the cross-correlation method, we looked for \ion{Ca}{i}, \ion{Cr}{i}, FeH, \ion{Fe}{i}, \ion{Fe}{ii}, H$_2$O, \ion{Li}{i}, \ion{Mg}{i},  \ion{Na}{i},  \ion{Ti}{i}, TiO, \ion{V}{i}, VO, and \ion{Y}{i}. Our results show no evidence  of any of these species in WASP-122b's atmosphere. The lack of significant detections can be explained by either the RM effect covering the regions where the atmospheric signal is expected and masking it, along with the low signal-to-noise ratio (S/N)  of the observations or the absence of the relevant species in its atmosphere.

}


%-----------------------------------------------------------%

\maketitle

%%%%%%%%%%%%%% INTRODUCTION %%%%%%%%%%%%%%
\section{Introduction}
High-resolution spectroscopic observations are powerful tools for characterizing exoplanetary atmospheres. Thanks to different radial velocities, we are able to differentiate the signal coming from the Earth's atmosphere, the host star, and the atmosphere of the exoplanet. Hot Jupiters (HJ, $T\rm{_{eq}}$<2000 K) and ultra-hot Jupiters (UHJ, $T\rm{_{eq}}$>2000 K), namely, gas giant planets with short orbital periods that are close to their host stars with hot and extended atmospheres, are the most suitable exoplanets for these studies. It is believed that due to their tidally locked nature leading to a large temperature gradient within the atmosphere, the chemistry in the night and day regions of the planet can be different \citep{Arcangeli2018}.

Recent transmission spectroscopic studies of the atmospheres of hot Jupiters using ground-based telescopes show a variety of species in their atmospheres. The most comprehensive studies of HJ using high-resolution transmission spectroscopic observations have been performed on HD~189733b \citep{Salz2018He, alonso_floriano_2019_hd189733, 2019_cabot_hd189, 2019_brogi_line_hd189_hd209, 2020_guilluy_hd189, 2021_boucher_hd189,Cristo_CaRM_2022,Cristo_hd189_2023} and HD~209458b \citep{Snellen2010, 2019_brogi_line_hd189_hd209, AlonsoFloriano2019HD209, sanchez_lopez_2019_hd209458,Santos_hd209_2020,2021_Giacobbe_hd209,Casasayas_HD209_2021}. The planets show the presence of molecules like H$_2$O, CO, HCN, CH$_4$, and NH$_3$, along with \ion{He}{i} and \ion{Na}{i}.

High-resolution detections of atoms and molecules were presented in several other HJs: 51 Pegb \citep{birkby_2017_sysrem, 51peg_2019_co_water}, HD~102195b \citep{Gloria2019},  HD~149026b \citep{Ishizuka_2021_hd149026_Ti_Fe}, tau Boo Ab \citep{Brogi_tauboo, tauboo2014, Tauboo2022},  WASP-21b \citep{2020_chen_wasp-21_Na}, WASP-52b \citep{2020_chen_wasp_52_Na_K_H}, WASP-69b \citep{2018_nortman_wasp69_He, 2017_casasayas_wasp69_Na, 2021_Khalafinejad_wasp69_na}, WASP-77Ab \citep{2021_Line_CO_water_C_O}, WASP-127b \citep{2020_allart_wasp127_Na,W127_Boucher_2023,W127_Nortmann_2024}, and WASP-172b \citep{W172_Seidel_2023}. The list of detected species consists of \ion{C}{i}, CO, \ion{Fe}{i}, \ion{H}{i}, H$_2$O, \ion{He}{i},  \ion{K}{i}, \ion{Na}{i}, OH, and \ion{Ti}{i}.

%51 Peg b - H2O (HR IR), CO (HR IR)
%HD149026 Ti (HR vis), Fe (hr ir)
%HD 189733  H2O (hr ir), CO(hr ir) , HCN (hr ir), He (hr ir), He+ (hr ir)
%HD209458  H2O (hr ir), CO (hr ir), He (hr ir), CH4 (hr ir), HCN (hr ir), NH3 (hr ir),

%tau Boo A CO (hr ir), H2O (hr ir)
%WASP-21 Na (hr vis)
%WASP-52 Na, K, H (hr -vis)
%WASP-69 He (hr ir),  Na (hr vis)
%WASP-77 A CO (hr ir), H2O(hr ir), C(hr ir), O(hr ir)
%WASP-127 Na (hr vis)

Compared to hot Jupiters, due to the much higher equilibrium temperatures, in ultra-hot Jupiters we expect mostly neutral and ionized atoms, where most of the molecules are dissociated. For example, KELT-9b, the hottest UHJ known to date, shows the presence of \ion{Ca}{i}, \ion{Ca}{ii}, \ion{Cr}{i}, \ion{Cr}{ii}, \ion{Fe}{i}, \ion{Fe}{ii}, H, \ion{Mg}{ii}, the Mg triplet, \ion{Ni}{i}, \ion{O}{i}, \ion{Na}{i}, \ion{Sc}{ii}, \ion{Sr}{ii}, \ion{Tb}{ii},\ion{Ti}{ii}, \ion{Y}{ii}, and evidence of \ion{Co}{i} and \ion{Sr}{ii} \citep{YanKELT9,  Hoeijmakers_2018_kelt9, Cauley_2019_kelt9, Hoeijmakers_2019_kelt9, yan-2019-calcium-kelt9-wasp33, Pino_2020_Kelt:9, 2020_Wyttenbach_kelt9_H,2021_Borsa_kelt9_O}.

%Other studies using transmission spectroscopic observations of UHJs have been performed on: HAT-P-70b \citep{2022_Bello-Arufe_hat_p_70}, MASCARA-2b/KELT-20b \citep{Casasayas2018, Casasayas2019, Stangret_2020_MASCARA-2, Nugroho2020_KELT20, Hoeijmakers_mascara2}, WASP-12b \citep{Jensen_2018_wasp12}, WASP-19b \citep{sedaghati_2017, K_19_2021}, WASP-33b \citep{yan-2019-calcium-kelt9-wasp33, Nugroho_2017_wasp33, Yan_2021_wasp33},  WASP-76b \citep{seidel-2019-wasp-76,Seidel_2021_w76, Ehrenreich_wasp76, 2021_Kesseli_Snellen_wasp76, 2021_Casasayas_wasp76,2021_Deibert_wasp76, 2021_Landman_wasp76,Tabernero_w76_2021, 2022_Kawauchi_wasp76, 2022_sanch_lopez_wasp76,2022_Kesseli_wasp76,Silva_w76_w121_2022}, WASP-121b \citep{2020HoeijmakersHEARTS,2020Cabot_wasp-121, 2020BenYami_wasp121, 2021_Borsa_wasp121, 2021_Merritt_wasp121,Seidel_w121_2023,Silva_w76_w121_2022}, and WASP-189b \citep{Yan_2020_wasp-189, 2021_Stangret_6uhj, 2022_Prinoth_wasp189}. 

In addition to the atmospheric composition, precise radial velocity measurements and analysis of the spectral line deformation, the so-called Rossiter-McLaughlin effect \citep{Holt1893,Schlesinger1910,Rossiter1924,McLaughlin1924} allows us to measure the sky-projected spin-orbit angle ($\lambda$). This allows us to gain valuable insight into the architecture of the system and to differentiate between different models of planetary migration (cite). It has been observed that planets orbiting stars with the $T_{eff}$ <  6250 K (\citealt{2010_winn}) are located on aligned orbits, while the planets orbiting warmer stars to do follow this trend. This behavior is explained by the Kraft break (\citealt{Kraft_break_1967}), where the stars with $T_{eff}$ <  6250 K experience angular momentum loss, which affects the spin-orbit alignment. The measurements of the obliquity of the system were successfully obtained for $\sim$ 200 transiting exoplanets \citep[i.e.,][]{Winn_obl_hd189,Esposito_2017_obl,Addison2018,Oshagh2018,Wang2018,Casasayas2020,Palle2020,Bourrier2021,Stangret2021}.  

These studies are possible thanks to the new generation of high-resolution spectrographs such as Echelle Spectrograph for Rocky Exoplanets and Stable Spectroscopic Observations (ESPRESSO, \citealt{espresso_pepe_2021}).
%\LEt{ Please set the full name first and the acronym in parentheses with the ref.***}(Echelle Spectrograph for Rocky Exoplanets and Stable Spectroscopic Observations, \citealt{espresso_pepe_2021}).
ESPRESSO is a fiber-fed échelle high-resolution spectrograph ($\Re \sim$ 140,000) mounted on the Very Large Telescope (VLT), in Cerro Paranal, Chile. The spectrograph covers the wavelength from 380 to 788~nm. It can be used on any of the four Unit Telescopes (UTs), as well as on four UTs simultaneously. 

\begin{figure}[]
\centering
\includegraphics[width=0.48\textwidth]{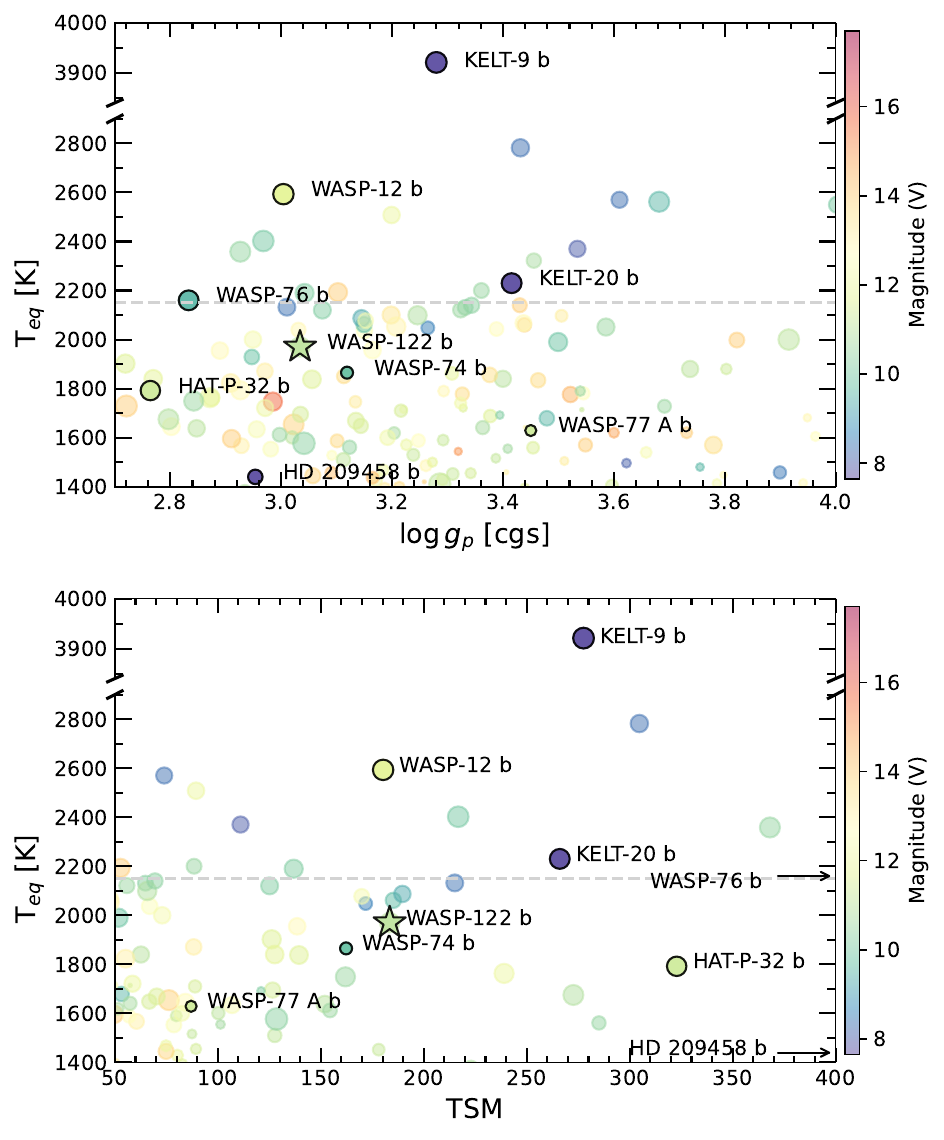}
\caption{Context of WASP-122b, marked with a star symbol, with respect to all known planets with $T_{eq}$> 1400K and $R_p$> 0.6 $R_J$. The equilibrium temperature is plotted on the vertical axis and surface gravity (top panel), and the transmission spectroscopy metrics (bottom panel) are plotted on the horizontal axis. The gray horizontal line in the temperature where we differentiate the HJ and UHJ according to \citet{2021_Stangret_6uhj} at 2150 K. For both figures, the size of the marker indicates the planetary radius and the V-band magnitude of the host star is color-coded.}
\label{fig:logg_TSM}
\end{figure}

In this work, we present studies of the architecture and atmosphere of hot Jupiter WASP-122b \citep{wasp-122_2016_Turner} also known as KELT-14b \citep{kelt-14_2016_Rodrigez},  using the ESPRESSO spectrograph in single UT mode (singleHR21 -high-resolution mode with 2x1 binning). The planet orbits a G2 star near the main sequence turnoff ($T_{\rm eff}$ = 5802~K, $M_{\star}$ = 1.178~$\rm{M_{\odot}}$,  $R_{\star}$ =  1.368 $\rm{R_{\odot}}$) with a period of 1.71 days. WASP-122b is an inflated gas giant planet with a mass of 1.196 $\rm{M_{J}}$ and a radius of 1.52 $\rm{R_{J}}$. Due to its close distance to the host star, WASP-122b has a high equilibrium temperature of 1904 $\pm$ 54 K. Its temperature, as well as the brightness of the star ($V=$ 11 mag), make this planet suitable for studying the chemical properties of its atmosphere using high-resolution ground-based spectrographs. All parameters of the star, the planet, and the system can be found in Table \ref{tab:params}. In Fig. \ref{fig:logg_TSM}, we present the context of WASP-122b with respect to all known planets with $T_{eq}$ > 1400 K and $R_p$ > 0.6 $\rm{R_{J}}$, depending on their equilibrium temperature versus surface gravity and transmission spectroscopy metrics (TSM, \citealt{Kempton2018}).  A TSM of $\sim$ 180 puts WASP-122b in the region where atmospheric chemistry is possible, but  its study is not straightforward. 

\begin{table}[h!]
\small\small
\centering
\caption{Physical and orbital parameters of the WASP-122 system.}
\begin{tabular}{lr}
\hline \hline
\\[-1em]
 Parameter  & Value \\ \hline
 \\[-1em]
  \multicolumn{2}{c}{\dotfill\it Stellar parameters \dotfill}\\\noalign{\smallskip}
 %\multicolumn{3}{c}{\it Stellar parameters}\\\noalign{\smallskip}
   \\[-1em]
\quad  $T_{\rm eff}$ [K] & $5802^{+95}_{-92}$  \\
  \\[-1em]
\quad  $\log g$ [cgs]& $4.234^{+0.045}_{-0.041}$ \\
  \\[-1em]
\quad  [Fe/H] & $0.326^{+0.091}_{-0.089}$ \\
  \\[-1em]
\quad $M_{\star}$ [$\rm{M_{\odot}}$]& $1.178^{+0.052}_{-0.066} $  \\
  \\[-1em]
\quad $R_{\star}$ [$\rm{R_{\odot}}$]& $1.368^{+0.078}_{-0.077}$  \\
  \\[-1em]
 %\quad  $v\sin i_{\star}$$^a$ [km\,s$^{-1}$]& $0.90^{+0.09}_{-0.12}$ \\
  \quad  $v\sin i_{\star}$$^a$ [km\,s$^{-1}$]& 1.84 $\pm$ 0.08 \\
  \\[-1em]
  \multicolumn{2}{c}{\dotfill\it Planet parameters \dotfill}\\\noalign{\smallskip}
% \multicolumn{3}{c}{\it Planet parameters}\\\noalign{\smallskip}
   \\[-1em]
 \quad $M_{\rm p}$ [$\rm{M_{J}}$]&  $1.196 \pm 0.072$ \\
  \\[-1em]
 \quad $R_{\rm p}$ [$\rm{R_{J}}$]& $1.52^{+0.12}_{-0.11}$  \\
 \\[-1em]
  \quad $T_{\rm eq}$ [$\rm K$]& $1904\pm 54$  \\
 \\[-1em]
\quad $K_{\rm p}$$^a$ [km\,s$^{-1}$]& 185 $\pm$ 4\\
  \\[-1em]
  \multicolumn{2}{c}{\dotfill\it Transit parameters \dotfill}\\\noalign{\smallskip}
% \multicolumn{3}{c}{\it Transit parameters}\\\noalign{\smallskip}
   \\[-1em]
 \quad $T_{\rm 0}$$^b$ [BJD$_{\rm TDB}$] & 2457872.52231 $\pm$ 0.00015   \\
  \\[-1em]
\\[-1em]
 \quad $P$ $^b$ [days] & 1.71005344 $\pm$ 0.00000032 \\
  \\[-1em]
 \quad $T_{14}$$^c$ [days] & $0.0889^{+0.0025}_{-0.0026}$  \\
 \\[-1em]
  \multicolumn{2}{c}{\dotfill\it System parameters \dotfill}\\\noalign{\smallskip}
% \multicolumn{3}{c}{\it System parameters}\\\noalign{\smallskip}
   \\[-1em]
  \quad $a/R_{\star}$& $4.64^{+0.25}_{-0.22}$ \\
 \\[-1em]
 \quad $R_{\rm p}/R_{\star}$& $0.1143^{+0.0029}_{-0.0026}$\\
 \\[-1em]
 \quad $i_{\rm p}$ [deg]& $79.67^{+0.80}_{-0.77}$  \\
 \\[-1em]
 %\quad $e$& - & \\
%\\[-1em]
%\quad $K_{\star}$ [m\,s$^{-1}$]& $179.8 \pm 9.0$  & \citet{kelt-14_2016_Rodrigez} \\
%  \\[-1em]
% \quad $\lambda$$^a$ [deg]& $-1.71^{+2.72}_{-2.31}$   \\
   
  \quad $\lambda$$^a$ [deg]& $0.09^{+0.88}_{-0.90}$   \\
   \\[-1em]
%\quad $v_{\rm sys}$ [km\,s$^{-1}$]&   \\
%\\[-1em]
\lasthline
\end{tabular}
\tablefoot{The physical and orbital parameters for the WASP-122 system were adopted from \citet{kelt-14_2016_Rodrigez}, except for \tablefoottext{a} {value retrieved in this work} and \tablefoottext{b} {adopted from \citet{Edwards_2021}.} \tablefoottext{c} {$T_{14}$ is the total transit duration, between the first and fourth contacts of the transit.}}
\label{tab:params}
\end{table}

This paper is structured as follows. In Sect. \ref{sec:observations}, we describe the observations and data reduction, in Sect. \ref{sec:obliquity} we describe the obliquity measurements. Sections \ref{sec:lines} and \ref{sec:cc_analy} describe the transmission spectra analysis, focusing on single lines and using the cross-correlation method, respectively. The final results are discussed and concluded in Sect. \ref{sec:discussion_conlucison}.

%%%%%%%%%%%%%% OBSERVATIONS %%%%%%%%%%%%%%.
\section{Observations and data reduction} \label{sec:observations}

\begin{figure}[]
\centering
\includegraphics[width=0.45\textwidth]{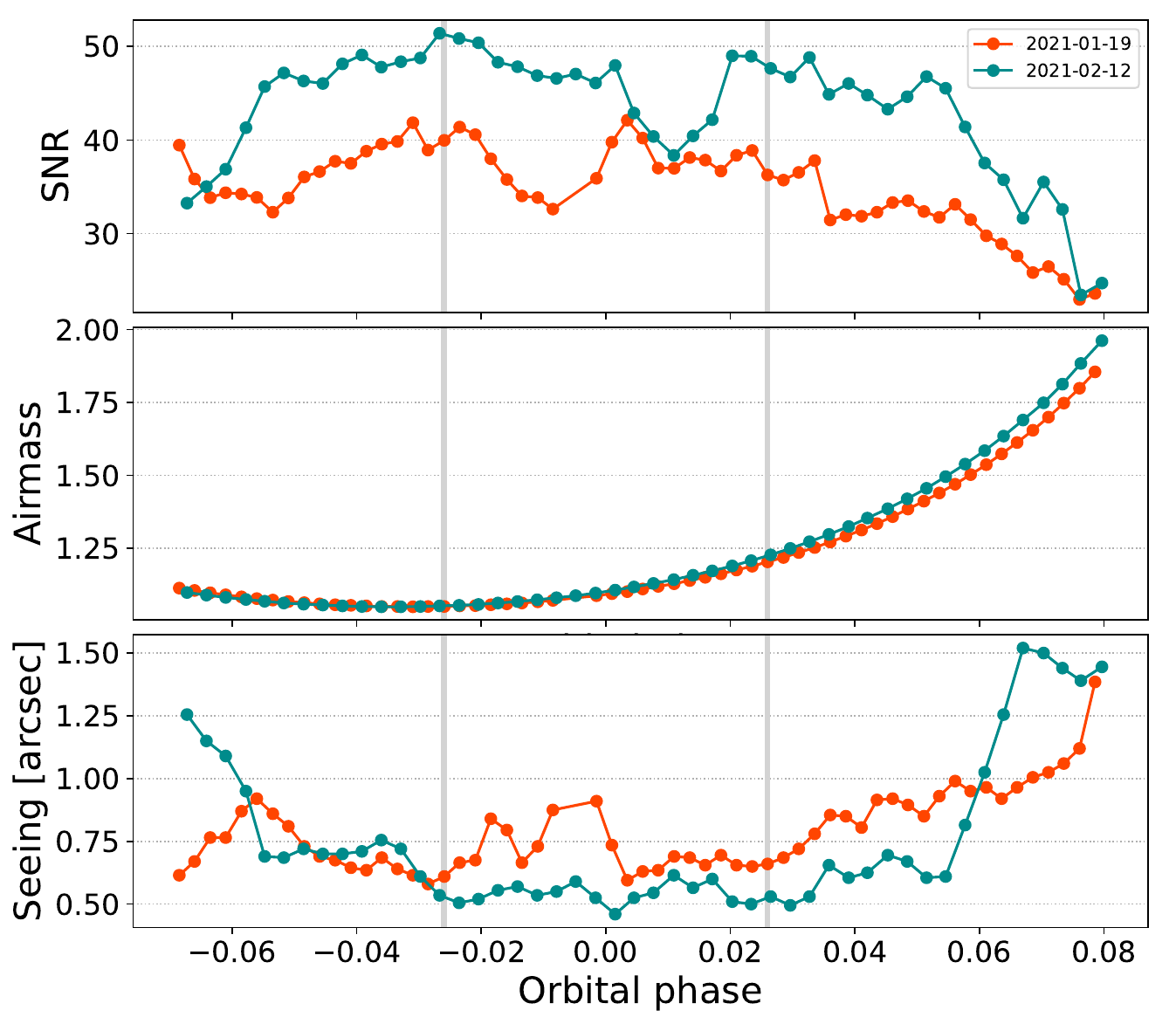}
\caption{S/N (top panel) in the \ion{Na}{i} echelle order (order 104), airmass (middle panel), and seeing (calculated as the mean seeing at the beginning and end of the exposure) evolution during the two nights of observations. Gray vertical lines represent the beginning (T1) and end (T4) of the transit. }
\label{fig:SNR}
\end{figure}

We observed two full transits of WASP-122b during the nights of 18 January 2021 and 12 February 2021 (hereafter referred to as Night 1 and Night 2), with the ESPRESSO spectrograph (Guaranteed Time Observation, program 106.21M2.004). During Night 1, we took 58 spectra with an exposure time of 300 s, 38 out-of-transit and 20 in-transit spectra, covering the planetary orbital phases $\phi$=-0.073 to $\phi$=0.073 with a mean signal-to-noise ratio (S/N) of 34.9 around 590 nm (physical echelle order 104). During Night 2, we took 48 spectra with an exposure time of 400 s, 31 out-of-transit and 17 in-transit spectra covering the planetary orbital phases $\phi$=-0.072 to $\phi$=0.074 with mean S/N of 43.4 around 590 nm. Observing log can be found in Table \ref{tab:obs}. The data were reduced using the Data Reduction Software (DRS) pipeline 2.2.8 and s1d sky-subtracted spectra were used in this work. The S/N and airmass evolution plots for each of the nights are shown in Fig. \ref{fig:SNR}.

\begin{table*}[]
\centering
\caption{Observing log of WASP-122b transit observations.}
\begin{tabular}{ccccccccc}
\hline\hline
Night & Date of   & Telescope     & Start & End &  Airmass\tablefoottext{a} &$T_\mathrm{exp}$ & $N_\mathrm{obs}$ & S/N\tablefoottext{b} \\
         & observation & & [UT] & [UT] & change & [s] &         &    \\ \hline
\\[-1em]
1 & 2021-01-19 & VLT-UT2 & 02:27 & 08:30 & 1.11-1.12-1.86 & 300 & 58 &  22.9-41.1 \\
\\[-1em]
2 & 2021-02-12 & VLT-UT3 &  01:04 &  07:06 & 1.10-1.13-1.96 & 400 & 48 & 23.4-51.4\\ 
\\[-1em]
\hline
\end{tabular}\\
\tablefoot{\tablefoottext{a}{Airmass change during the observations, where the values represent, airmass at the beginning of the nights, mid-observations, and at the end of the night, respectively.} \tablefoottext{b}{Minimum and maximum S/N for each night, calculated in Na I echelle order (physical order 104).}}
\label{tab:obs}
\end{table*}

%%%%%%%%%%%%%% Measurement of the obliquity %%%%%%%%%%%%%%
\section{Obliquity measurement} \label{sec:obliquity}
%1. We need plot: Teff x Obliquity x Planet Radius

%In order to measure the spin-orbit angle of the system we used the classical Rossiter-McLaughlin 

 The classical Rossiter-McLaughlin (RM) effect offers information about the planet-star radius ratio, the rotational velocity of the host star, impact parameter, and obliquity of the system \citep{TriaudReview2018}. In particular, the modeling of the RM effect has been widely used in the literature as a tool to measure the sky-projected spin-orbit angle $\lambda$ \citep[i.e.,][]{Addison2018,Oshagh2018,Wang2018,Casasayas2020,Palle2020,Bourrier2021}.

\begin{figure}[h!]
    \centering
    \includegraphics[width=0.45\textwidth]{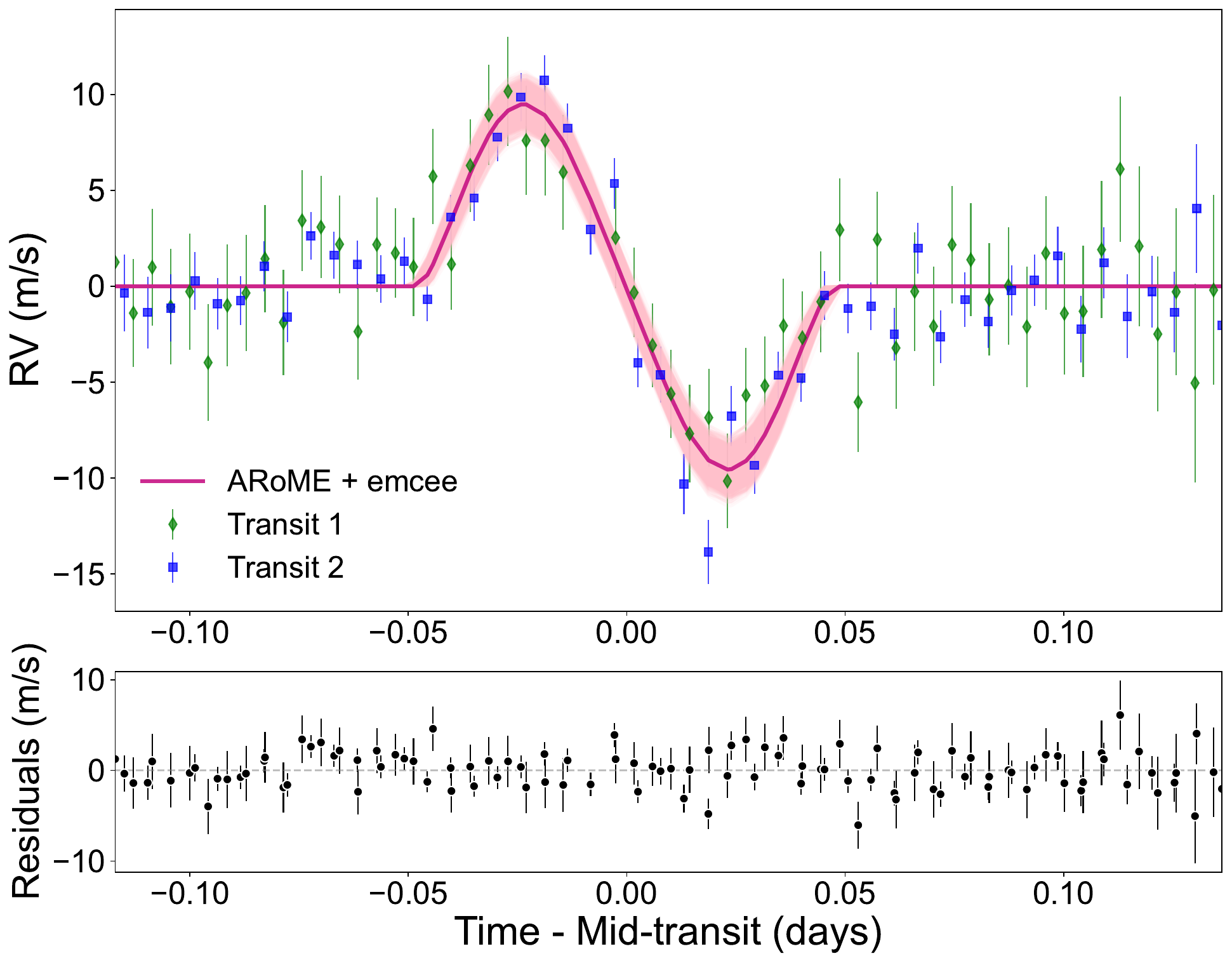}
    \caption{The Rossiter-McLaughlin signal from the Night 1 (green diamonds) and Night 2 (blue squares) RV time series (top). The solid lines show the best RM model resulting from the \texttt{ARoME} + \texttt{emcee} fitting (magenta) and models within 1$\sigma$ (light pink). Bottom panel shows the residuals.}
    \label{fig:RM_fitting}
\end{figure}

Here, we analyzed the RM signal present in the radial velocity (RV) measurements extracted by the DRS pipeline from both ESPRESSO datasets using G8 binary mask. For each of the relevant nights, we extracted the RM signal by subtracting a linear fit to the out-of-transit RVs. Subsequently, we combined both data sets by sorting them in orbital phase. In Fig.~\ref{fig:RM_fitting} we plot the measurements from each of the separate nights.% and for a better view binned measurements using 0.0025 binning in phase (around 6.15 min).  }

%{\color{red}We combined the data sets of both nights using a 7.2~min binning and extracted the RM signal by subtracting a linear fit to the out-of-transit RVs. Thereby, we used the combined data set for the RM analysis (Figure~\ref{fig:RM_fitting}).}

The approach used by the ESPRESSO DRS pipeline to measure the RVs is based on the fitting of a Gaussian to the cross-correlation function (CCF), which is a function obtained from the cross-correlation of a spectrum with a binary mask \citep[this technique is detailed in][]{Pepe2002}. For this reason, to model the RM signal, we used a Python implementation of \texttt{ARoME} \citep{Boue2013}, which is a code specifically designed to model the RV data extracted by the CCF approach. We also utilized \texttt{emcee} \citep{emcee} to perform a Markov chain Monte Carlo (MCMC) fitting procedure of the RM signal using the \texttt{ARoME} models. As free parameters, we took the mid-transit time ($t_{mid}$), the sky-projected spin-orbit angle ($\lambda$), and the projected stellar rotational velocity ($v\sin{i}$). We imposed large uniform priors on the free parameters: ($-0.02$~days $<T_{0}-t_{mid}<0.02$~days), ($-45$~deg $<\lambda<45$~deg), and (-30~km\,s$^{-1}$ $< v \sin{i} <$ 30~km\,s$^{-1}$). Throughout the fitting, we fixed the rest of the planetary and stellar parameters, including the limb darkening coefficients, which were fixed to the values obtained through \texttt{LDTk} \citep{Parviainen2015LDTk} estimations. By assuming quadratic limb-darkening law the limb-darkening coefficients were fixed to $q_1 =0.577 \pm 0.002$ and $q_2 = 0.113 \pm 0.003$.
% From ARoME paper: RM effect specially designed to model observations done by the Gaussian fit of a cross-correlation function (CCF) as in the routines performed by the HARPS team.

As a result of the fitting procedure, we determined the sky-projected spin-orbit angle to be $\lambda = 0.09^{+0.88}_{-0.90}$~deg. In Fig.~\ref{fig:RM_fitting}, we show the best RM model and the $1\sigma$ uncertainties from the fit, the fitted values are given in Table \ref{tab:RM_params}, and the corner plots from the fitting are shown in the \ref{fig:RM Corners and chains}.

\begin{figure}[h!]
    \centering
    \includegraphics[width=0.48\textwidth]{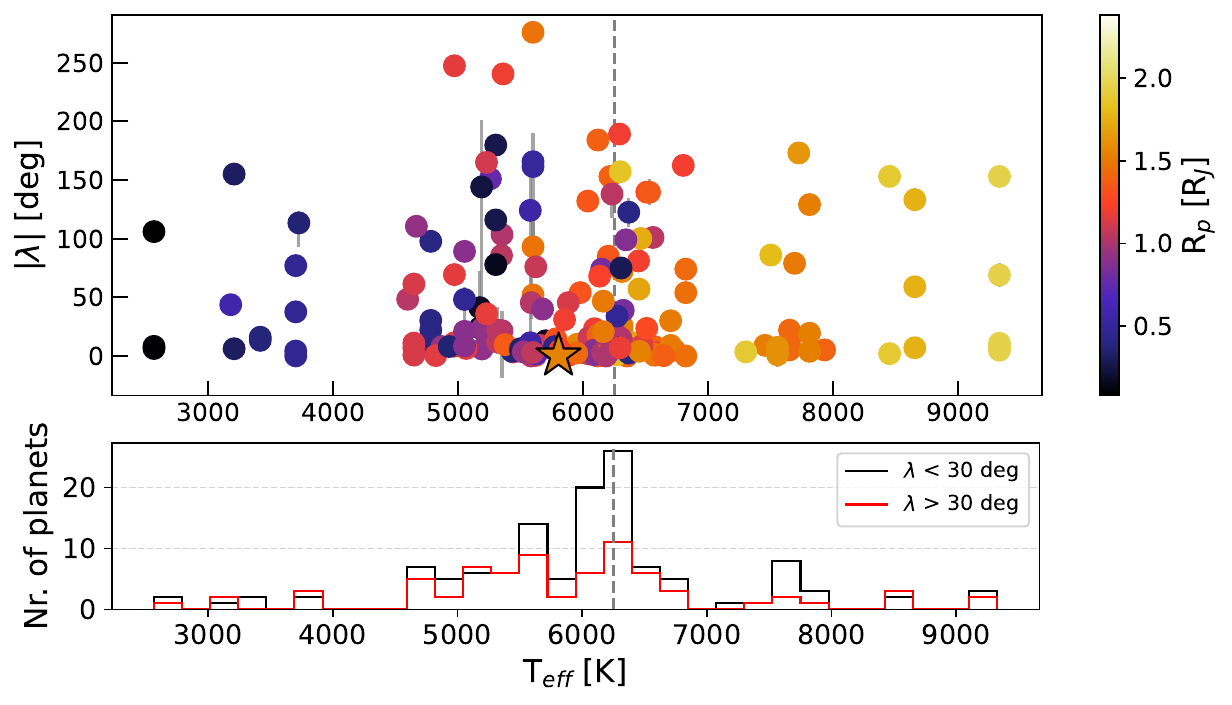}
    \caption{WASP-122b in the $T_{eff}-|\lambda$| plane compared to the rest of the confirmed planets with an obliquity measurement (top). The colors indicate the planetary radius. WASP-122b is marked with a "star" symbol. The sample of planets with an obliquity measurement was extracted from the TEPCat catalog \citep{Tepcat_2011}. In the case of planets with more than one measurement, we plot the mean of those values, with the error bar indicating the span between the maximum and minimum literature values.  Bottom panel: Histogram of number of planets with $\lambda$ > 30 deg and $\lambda$ < 30 deg, depending on the $T_{eff}$ of their main star. In both panels, vertical grey dashed lines indicate the $T_{eff}$ = 6250 K. }
    \label{fig:Obliquity_demographics}
\end{figure}

Thus, we found WASP-122b to be an aligned planet. In the top panel of Fig.~\ref{fig:Obliquity_demographics}, we compare WASP-122b to the rest of the known planets with an obliquity measurement in the  $T_\textrm{eff}-\lambda$ plane, indicating the size of each planet in color. WASP-122b follows the general alignment trend shown by planets orbiting stars with effective temperatures cooler than 6250~K, and is one of the largest planets in this population. This $T_{eff}$ is directly related to the Kraft Break (cite). Stars with $T_{eff}$ < 6250~K experience angular momentum loss due to a large convective zone, which directly influences the spin-orbit angle of the planets around them.

%\textbf{In bottom panel of Figure~\ref{fig:Obliquity_demographics}, we how the histogram xxxxx} 

\begin{table}[h!]
\small\small
\centering
\caption{Emcee + PyArome combined fitting results of the parameters.}
\begin{tabular}{l l}
\hline \hline
\\[-1em]
 Parameter  & Value \\ \hline
 \\[-1em]

 %\multicolumn{3}{c}{\it Stellar parameters}\\\noalign{\smallskip}
\\[-1em]
$T_{0} - t_{mid}$ [days] &  $0.0002 \pm 0.0007$  \\
\\[-1em]
$\lambda$ [deg] &  $0.09^{+0.88}_{-0.90}$  \\
\\[-1em]
$v \sin{i}$ [km\,s$^{-1}$] &  $1.84 \pm 0.08$   \\
\\[-1em]
\lasthline
\end{tabular}
\label{tab:RM_params}
\end{table}

%{\color{red}Add rest of fitting results of the parameters???} {\color{green}Monika's comment: yes, in table will be the best}
%Emcee + PyArome combined results: $t_{0} - Mid-transit = 0.0085^{+0.0012}_{-0.0012}~days$ $\lambda = 0.85^{+1.45}_{-1.36}~deg$ $vsini = 0.921^{+0.066}_{-0.070}~km/s$

%N1$t_{0} (MA) = 0.0098 \pm 0.0015~days$
%N1$\lambda = -1.5^{+2.3}_{-2.1}~deg$
%N1$vsini = 0.89^{+0.09}_{-0.10}~km/s$

%\section{The system velocity of KELT-14b}

%%%%%%%%%%%%%% ATMOSPHERIC CHARACTERIZATION %%%%%%%%%%%%%%

%\section{Reloaded RM} \label{sec:reloaded_RM}

%{\color{red}Vincent}

There is no rotation period available for the star WASP-122b in the literature, since the star is inactive \citealt{wasp-122_2016_Turner}. Thus, measuring the starspot rotational period using spectroscopic time series is extremely difficult, so the true obliquity angle of the planetary system cannot be determined. However, if the system is synchronized (expected given the mass and short orbital period of the planet, \citealt{kelt-14_2016_Rodrigez}) and both star and planet share the same rotation rate, we can estimate an obliquity angle of $\psi$ = $\pm$ 77.06 $\pm$ 0.86 deg.

\section{Transmission spectroscopy around single lines} \label{sec:lines}

Following the methodology presented in \citet{Wytt2015} and \citet{Casasayas2018}, and later applied to ESPRESSO data in \citet{Casasayas_HD209_2021}, we determined the transmission spectrum of WASP-122b around several atmospheric lines previously found in hot atmospheres; namely: the \ion{Na}{i} doublet ($\lambda$5891\r{A} and $\lambda$5897\r{A}), H$\alpha$, H$\beta$, and \ion{Li}{i} ($\lambda$6709\r{A}).

The extracted spectra were first telluric corrected using the {\tt Molecfit} code (\citealt{Molecfit1}, \citealt{Molecfit2}), and then shifted to the stellar rest frame, taking into account the barycentric correction, the systemic velocity, and the movements of the star around the center of mass of the system; the latter were corrected by fitting the linear function to all of the out-of-transit radial velocity measurements, assuming eccentricity equal to zero (\citealt{wasp-122_2016_Turner}, \citealt{kelt-14_2016_Rodrigez}). After normalizing the spectrum, the stellar signal was removed by dividing each spectrum by the master-out spectrum, calculated as the mean of all of the out-of-transit spectra. The resulting residuals were then shifted to the planetary rest frame using the formula:

\begin{equation}
    v_p(t, K_p)= K_p \sin{2 \pi \phi (t)}
    \label{formula:vp}
,\end{equation}

where $v_p$ is the radial velocity of the planet, $K_p$ is the semi-amplitude of the exoplanet radial velocity, and $\phi$ is the orbital phase of the planet. In this case, we assumed the $K_p$ to be 185 ~km\,s$^{-1}$, which is equal to the predicted $K_p$ value calculated using the orbital parameters from Table \ref{tab:params}. Finally, all in-transit residuals were combined to obtain the exoplanet's transmission spectrum. As shown in several previous works using the ESPRESSO spectrograph (e.g., \citealt{2020_allart_wasp127_Na}, \citealt{Tabernero_w76_2021}, \citealt{Casasayas_HD209_2021} ), the spectra are affected by two sinusoidal patterns created by the coude train optics, called ``wiggles,'' with periods of 30 $\AA$ and 1 $\AA$. Due to the low S/N of the observations, the 1 $\AA$ wiggles were not detected. The wiggles with period 30 $\AA$ were not clearly detected in the single spectra in the regions around the studied lines, but created the sinusoidal structure in the final transmission spectrum. We corrected this anomaly locally around the studied line by fitting a fourth-order polynomial function.

%, and normalized, and all of the out-of-transit spectra are combined to build a high SNR template spectrum. Then we compute the ratio between each spectrum and the master template to remove the stellar contribution. The resulting residual maps are shifted to the planetary rest frame, and combined to obtain the transmission spectrum of the exoplanet.

\begin{figure*}[h!]
    \centering
    \includegraphics[width=\textwidth]{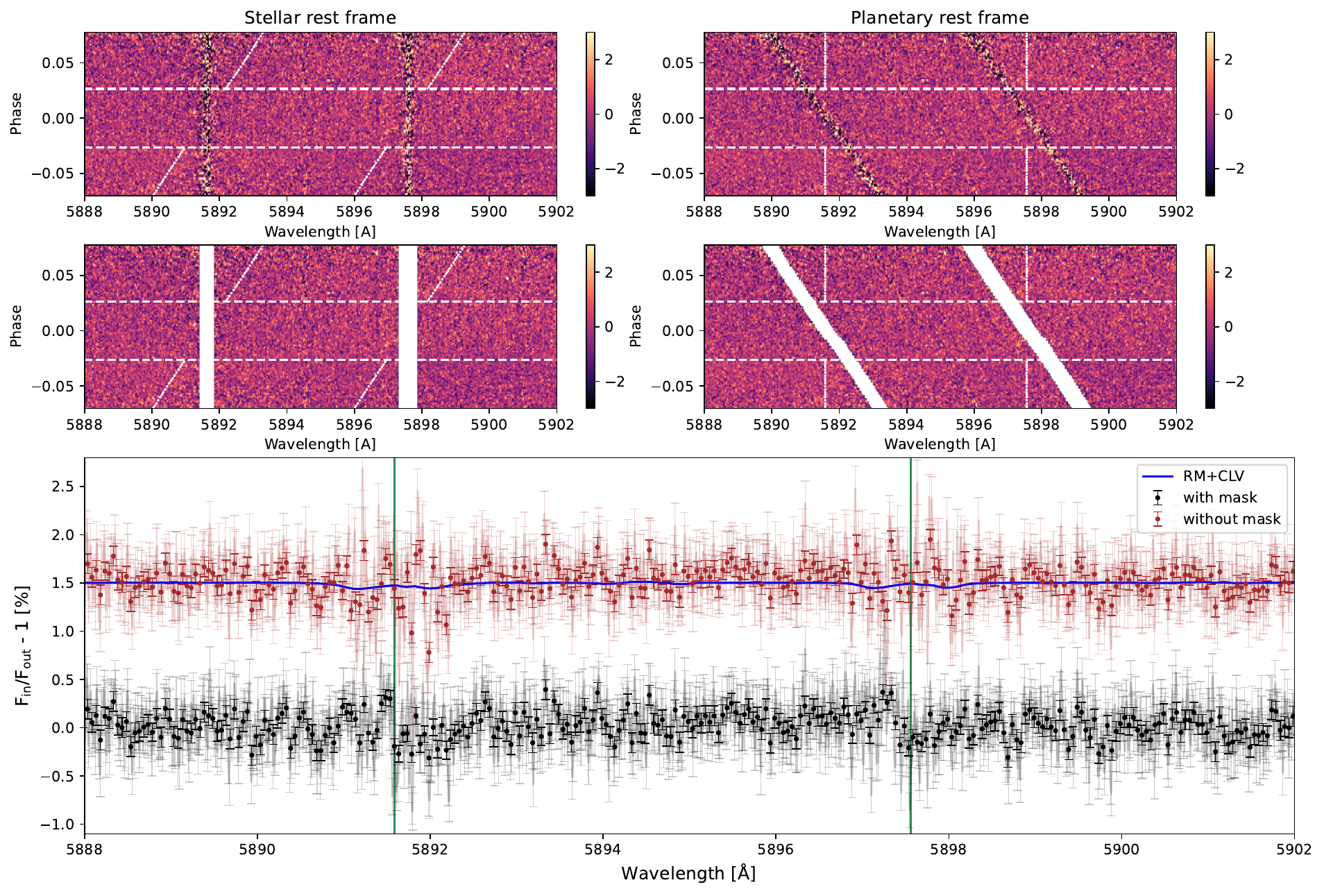}
    \caption{2D map of the individual transmission spectra around the \ion{Na}{i} doublet lines of WASP-122b (top) in the stellar (left) and planetary (right) rest frame. Middle panel shows the same as the top panels, but masking the stellar signal of the \ion{Na}{i} doublet lines (presented mask size $\pm$ 0.2 $\AA$). Bottom panel gives the transmission spectrum computed by combining the data between the first and fourth contacts of the transit, indicated in the top panel. The red plot was computed without masking the stellar \ion{Na}{i} lines, and the black plot with masking of these lines. Blue line represents the RM and CLV effects model. We have chosen not to show the TS after removal since the effect is marginal. The green vertical lines indicate the laboratory position of the \ion{Na}{i} doublet lines.}
    \label{fig:transmission_Na}
\end{figure*}

After exploring the different spectral lines, we found no evidence of planetary absorption in any of them. In Fig.~\ref{fig:transmission_Na} we show the final results in the spectral region containing the \ion{Na}{i} doublet. The top two panels show two-dimensional maps of the individual transmission spectra around the \ion{Na}{i} doublet in the stellar and planetary rest frame, respectively. In the middle panels, we present similar maps as those in the top panels, but we have masked the regions $\pm$ 0.2 $\AA$ from the center of the lines affected by the residuals of the stellar \ion{Na}{i} doublet. Calculations were repeated by masking the regions around the studied lines from $\pm$ 0.1 to $\pm$ 0.5 $\AA$ with similar results.  The final transmission spectrum is shown in the bottom panel, without (red) and with (black) masking the stellar signals. The presented transmission spectrum shows no statistically significant features. Very similar results are found for the rest of the studied lines shown in Fig. \ref{fig:transmission_Ha} for H$\alpha$, Fig. \ref{fig:transmission_Hb} for H$\beta$, and \ref{fig:transmission_Li} for \ion{Li}{I}.

%The maps are given in the stellar rest frame, and it is easy to spot that the residuals observed at the doublet position are tilted and extend to the observations before and after transit times. Both things are clear indications that the residuals are stellar nature and not due to a planetary absorption.  In Figure~\ref{fig:transmission} we also show the derived transmission spectrum over the same spectral regions, showing no statistically significant features. Very similar results (not shown) are found for the rest of the lines.

\section{Survey of atomic and molecular species using the cross-correlation technique} \label{sec:cc_analy}

To study the chemical atmospheric composition of WASP-122b, we applied the cross-correlation (CC) method to the data using a range of atoms and molecules:  \ion{Ca}{i}, \ion{Cr}{i}, FeH, \ion{Fe}{i}, \ion{Fe}{ii}, H$_2$O,\ion{K}{i}, \ion{Mg}{i},  \ion{Na}{i},  \ion{Ti}{i}, TiO, \ion{V}{i}, VO, and \ion{Y}{i}. All the models were calculated with the petitRADTRANS code \citep{molliere_2019_petiRADTRANS} assuming solar abundances from \citet{Asplund2009SunAbund}, equilibrium chemistry, and two different temperatures $T_{eq}$= 1904 K ($T_{eq}$ of the planet) and 4000 K. We tested models for two temperatures because observed regions can have much higher temperatures than $T_{eq}$. It is possible that spectral lines for the models created for lower temperatures may not exist, which would result in the species not being detected using the CC method. Due to the absence or weakness of spectral lines for \ion{Fe}{ii} and \ion{Mg}{i} for $T_{eq}$= 1904 K, we decided not to show the results for these cases. In the case of TiO, we created two different models using the line lists provided by Exomol \citep{Exomol_2019} and Plez \citep{Plez_1998}, both available in the petitRADTRANS code. Finally,  the models were convolved to the resolution of the ESPRESSO spectrograph ($\Re=$140,000). We did not convolve the models to the rotational period of the exoplanet or the exposure time because the difference in cross-correlation is insignificant and does not affect the final detection or non-detections. The models and information about the line list can be found in Fig. \ref{fig:atoms_models}.

We reduced the data following \citet{2021_Stangret_6uhj}.  In the first steps, we repeated the procedures from the analysis of the single lines, where the telluric lines were corrected with the {\tt Molecfit} code, and the spectra were moved to the stellar rest frame. In the next step, we divided each spectrum into 10000-pixel orders, at each wavelength, excluding the pixels that were divided from fitted quadratic polynomials more than 5$\sigma$, which were related to the remaining hot and cold pixels. Each of the orders was normalized by fitting a quadratic polynomial. The emission lines larger than 5$\%$ were masked. Finally, the stellar signal was removed by dividing each spectrum by the master-out spectrum, calculated as the mean of all of the out-of-transit spectra. Before the CC, we combined each order into one spectrum, masking the wavelengths < 4500 \r{A}, where the S/N was low, and the regions between 6850-6950 \r{A} and 7550-7700 \r{A}, which are strongly affected by telluric lines.

Afterward, the residuals were cross-correlated with atmospheric models for radial velocities in the range of $\pm300$~km\,s$^{-1}$ with a step of $0.5$~km\,s$^{-1}$, which corresponds to the pixel size of ESPRESSO. We removed a fourth-order polynomial from each of the CC residuals in time, which brought the CC values to a similar level. This procedure effectively removed the broadband structures, such as big wiggles, without affecting the possible detection of the sinusoidal signal, which is much wider (>200 km~s$^{-1}$) than the expected planetary signal. The residual maps for each of the species are presented in the left column of Fig. \ref{fig:CC_1900_1} and Fig. \ref{fig:CC_1900_2} for 1904~K and in Fig \ref{fig:CC_4000_1} and \ref{fig:CC_4000_2} for 4000~K. In the next step, we shifted the maps to the planetary rest frame using Formula \ref{formula:vp}. In the case of the cross-correlation method, we assumed that $K_p$ is unknown and calculated $v_p$ for a range of $K_p$ values from 0 to 300 ~km\,s$^{-1}$, in steps of 1 ~km\,s$^{-1}$. Afterward, in order to retrieve higher S/N, we co-added the in-transit (including ingress and egress) cross-correlation residuals. In the final step, we calculate the significance of the results by calculating the S/N by dividing the CC values by the standard deviation calculated around -150~km\,s$^{-1}$ to -50 ~km\,s$^{-1}$ and 50 ~km\,s$^{-1}$ to 150~km\,s$^{-1}$

%cc values were co-added  %{\color{red}\citet{alonso_floriano_2018_hd189733}, and \citet{sanchez_lopez_2019_hd209458}, the in-transit cross-correlation values (excluding ingress and egress) for each $K_p$ independently were co-added, and we checked the significance of the signal by calculating its S/N by  dividing the result by the standard deviation calculated far from the position where we expect the signal (-150~km\,s$^{-1}$ to -50 ~km\,s$^{-1}$ and 50 ~km\,s$^{-1}$ to 150~km\,s$^{-1}$).}

\subsection{Rossiter-McLaughlin effect modelling and correction } \label{sec:RMcorr}

\begin{figure*}[h]
\centering
\includegraphics[width=0.9\textwidth]{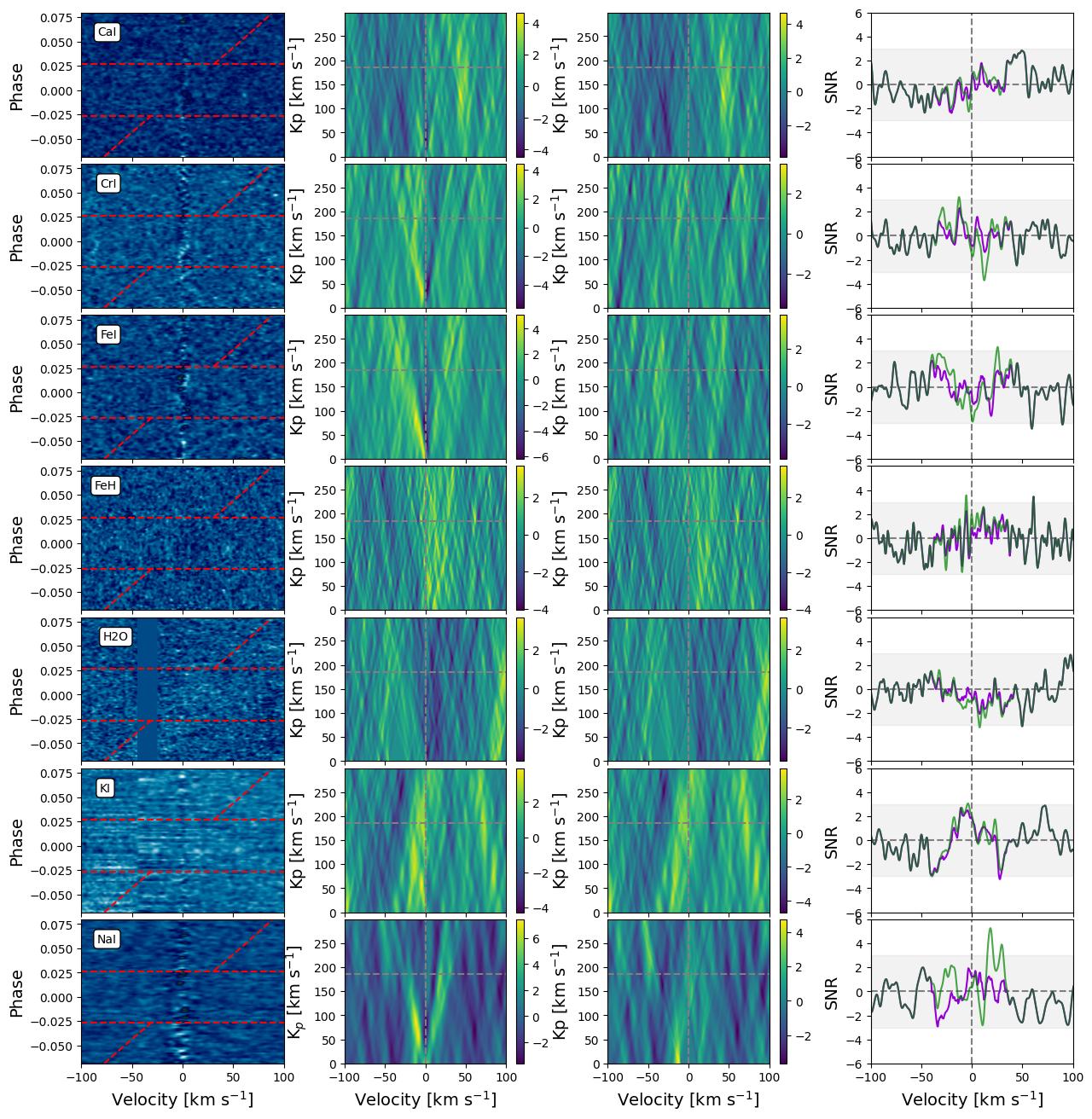}
\caption{ Cross-correlation results for WASP-122b for \ion{Ca}{i}, \ion{Cr}{i}, \ion{Fe}{i}, FeH, H$_2$O, \ion{K}{I}, and \ion{Na}{i}. For each of the species, we present the following plots. First column: Residual map before masking the RM effect, where the red horizontal lines represent the beginning and end of the transit and the red tilted line represents the expected velocity of the planet. Second column: $K_p$ map calculated for a range of 0 to 300 km\,s$^{-1}$, the theoretical $K_p$ of the planets is represented by a gray horizontal line. Expected signal from the planet should be detected at $K_p=$185 km\,s$^{-1}$ and radial velocity 0 km\,s$^{-1}$. Third column: Same as the second column but after masking the RM effect. Fourth column: S/N plot for theoretical $K_p$ value. The green plot is the raw S/N plot and the purple plot is S/N after masking the regions where the RM effect appears. The horizontal gray dashed line represents S/N = 0, and the vertical gray dashed line represents the radial velocity of 0 km/s, which is the expected position of the CC peak. The gray regions represent S/N between -3 and 3. }
\label{fig:CC_1900_1}
\end{figure*}

Although WASP-122 is not a fast-rotating star, due to the high quality of the data we were able to detect the Rossiter-McLaughlin (RM) effect in some of the cross-correlation residuals. As shown in the CC residual map especially for \ion{Fe}{i} in Fig. \ref{fig:CC_1900_1}, the RM effect covers some part of velocities, where we expect the signal coming from the planet and, in consequence, in the final $K_p$ maps and S/N plots, it can block or mimic the possible planetary signal. Removing this effect is a crucial step in the analysis. We applied two different methods to deal with this signal. 

In the first case, we masked the region in the stellar rest frame, where the RM appears - a region between -10 and 10 km\,s$^{-1}$. We note that for all pixels in this region, the value is assumed to be 0 since the CC values are co-added in the next step and removing them completely was not necessary.) Then we followed the steps described in Sect. \ref{sec:cc_analy} by co-adding the in-transit data and calculating the S/N plots. We masked the same regions for all of the studied species, even for the species where the RM effect was not detected by eye. The S/N plots after masking are shown by the purple line in the right columns of Figs. \ref{fig:CC_1900_1} and \ref{fig:CC_1900_2} for 1900 K, and Figs. \ref{fig:CC_4000_1} and \ref{fig:CC_4000_2} for 4000 K.

In the second case, we modeled the RM effect together with CLV \citet{Yan_CLV_2015} and \citet{YanKELT9}. First we created a high-resolution stellar spectrum model using the \texttt{Turbospectrum2019} code (\citealt{Turbospectrum2019}), together with  VALD line list assuming stellar parameters from Table \ref{tab:params} and PHOENIX \citep{Phoenix_2013} stellar model for $T_{\rm eff}=5750$ K, $\log{g}=4$, and $[Fe/H]=0.5$. 
%In the second case, we modeled the RM effect following \citet{Yan_CLV_2015} and \citet{YanKELT9}. Using the Gottingen Spectral Library and PHOENIX \citep{Phoenix_2013} we created a high-resolution stellar spectrum model assuming $T_{\rm eff}=5800$ K, $\log{g}=4$, and $[Fe/H]=0.5$. 
The models account for the fact that at each orbital phase, the planet covers a different region of the star with different limb-darkening and Doppler effect. Next, we divided the modeled spectrum by the stellar spectra computed outside the transit. This ensured that the stellar spectra from all regions of the star were included in the final spectra. The remaining signal is the RM effect. The modeled RM was convolved to the resolution of the spectrograph and cross-correlated with the atomic and molecular models described in Sect. \ref{sec:cc_analy}. In addition, the models were scaled by the S/N of each exposure around \ion{Na}{i} doublet.

\begin{figure}[h!]
\centering
\includegraphics[width=0.45\textwidth]{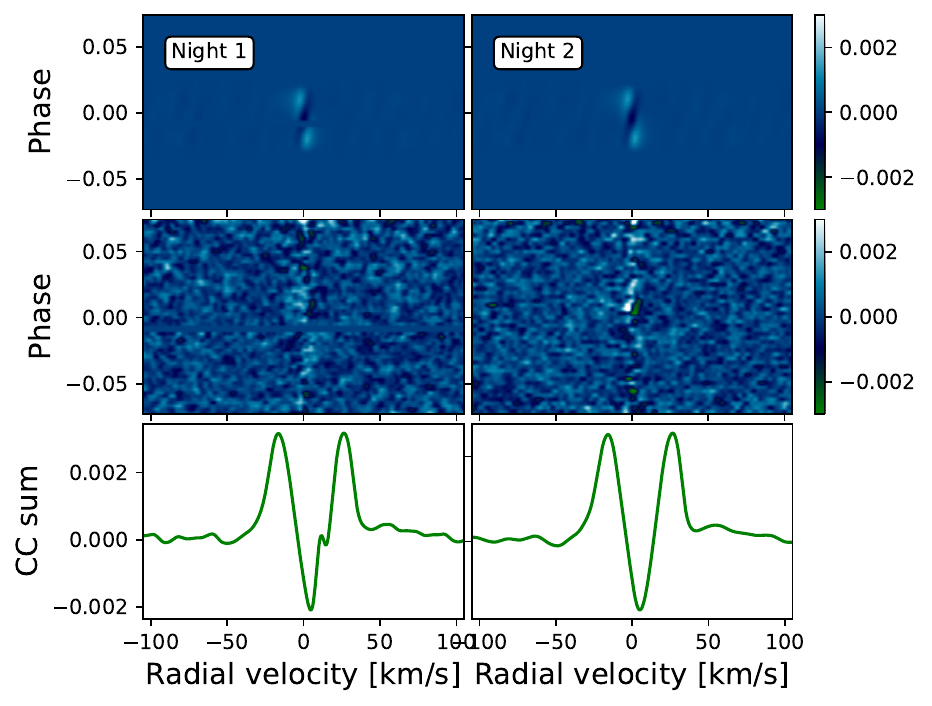}
\caption{Modeled RM+CLV effects for \ion{Fe}{i} for Night 1 and Night 2. In the top plots, we present the modeled RM+CLV effects after cross-correlation with the model of \ion{Fe}{i}. In the middle plots, we show the cross-correlation residual plots for Night 1 and Night 2 separately in order to compare the shape of the modeled RM+CLV effects. In the bottom panels, we present the CC, after summing the in-transit data of the RM effect for $K_p=$ 185 km\,s$^{-1}$. The feature in the RM+CLV model of Night 1 is due to a gap in the data during the observations.}
\label{fig:models_rm}
\end{figure}

As an example, in Fig. \ref{fig:models_rm} we present the final computed RM model for \ion{Fe}{i}. Since the S/N of the observations was not stable during the observations and WASP-122 is a slowly rotating star, the RM effect, in some of the spectra, was not detected or detected with low strength (see CC residual plots in Fig. \ref{fig:models_rm}). The final removal will indeed correct the RM effect in the spectra with the higher S/N but will add an additional signal in the spectra where the RM effect was not detected since the effects were not scaled by the S/N of the observations. In this case, we decided not to proceed with further analysis and did not apply this correction.

\subsection{Cross-correlation results}
 
\begin{figure*}[h]
\centering
\includegraphics[width=0.9\textwidth]{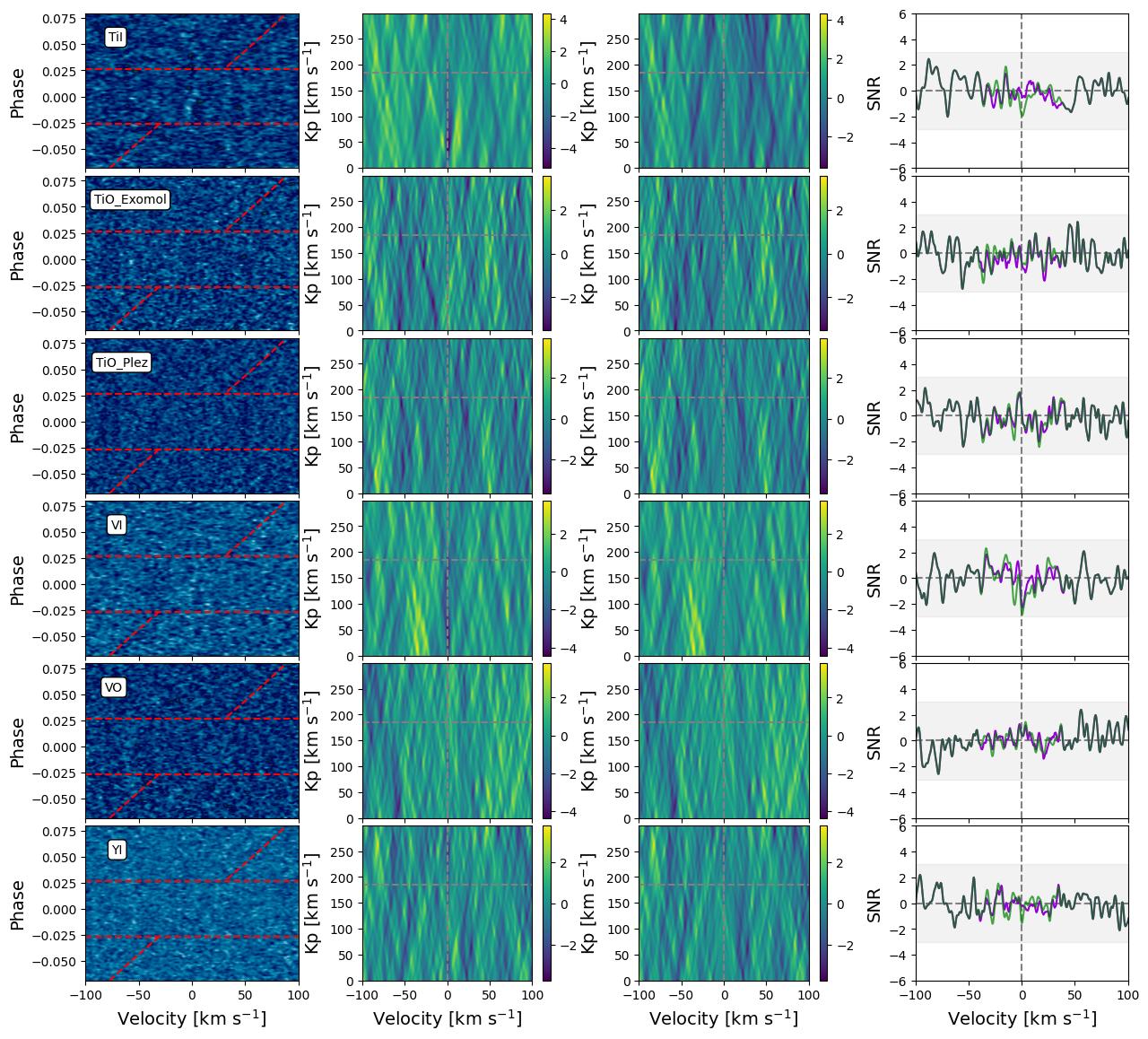}
\caption{ Same as Fig. \ref{fig:CC_1900_1} but for \ion{Ti}{i}, TiO (Exomol), TiO (Plez), \ion{V}{i}, VO, and \ion{Y}{i}.}
\label{fig:CC_1900_2}
\end{figure*}

The results of the cross-correlation analysis are shown in the Figs. \ref{fig:CC_1900_1} and \ref{fig:CC_1900_2}. In the left panels, we present the cross-correlation residual maps for each of the studied species, without masking the RM effect. In the case of the H$_2$O, we decided to mask the regions where the atmospheric water was detected and was not sufficiently corrected by the {\tt Molecfit} code. In the second column, we show the significance maps ($K_p$ maps), for the case without masking the RM effect. In the last column, we present S/N plots for the literature $K_p$. In the S/N plots, due to the fact that in some of the results we detected the RM effect (which partially lies in the regions where we would expect the signal from the planet), we show two plots. One represents the results without any correction and the second illustrates the results after we masked the region where the RM affects the data. We present the plots with masks for all the species studied, including those where the RM is not visible by eye. In all cases, we did not detect any significant signal coming from the planetary atmosphere.
We present the masked plots for all the species studied, including those where the RM is not visible.

\subsection{Measurement of upper limits - Injection of the signal \label{sec:uplim}}

To measure the upper limits of the non-detections, in the case of transmission spectroscopy studies around single lines, we calculated the one-pixel dispersion of the detection. The calculation was based on the standard deviation of the transmission spectrum away from the studied line for a band of 10 \r{A}. We assumed the upper limit to be 3$\sigma$. The calculated upper limits for the corresponding lines can be found in Table \ref{tab:Upper_sl}. These values represent the minimum absorption level that can be detected using dataset from this work.

\begin{table}[h!]
\small\small
\centering
\caption{Upper limit of 3 $\sigma$  for the detection of H$\alpha$, H$\beta$, Na D1 \& D2, and \ion{Li}{i}.}
\begin{tabular}{l l l}
\hline \hline
\\[-1em]
 Species  & 3 $\sigma$ [\%]  & \\ \hline
 \\[-1em]

 %\multicolumn{3}{c}{\it Stellar parameters}\\\noalign{\smallskip}
\\[-1em]
H$\alpha$ &  0.31  &   \\
\\[-1em]
H$\beta$ &  0.48 &    \\
\\[-1em]
Na D1 \& D2 & 0.69   &  \\
\\[-1em]
\ion{Li}{i} & 0.30   &  \\
\\[-1em]
\lasthline
\end{tabular}
\label{tab:Upper_sl}
\end{table}

Following \citet{2020HoeijmakersHEARTS}, \citet{Allart2017,2020_allart_wasp127_Na}, and \citet{Casasayas_Barris_mascara_1_2022_ESPRESSO} in order to calculate the upper limits of the detection using the cross-correlation method we injected synthetic atmospheric signal into the observational data at 200 ~km\,s$^{-1}$, away from the expected signal from the planetary atmosphere. In this step, the lines in the models were broadened considering the resolution of the spectrograph using \texttt{instrBroadGaussFast} routine from \texttt{PyAstronomy} \citep{Czesla_2019_pyastronomy}, the rotation of the tidally locked planet ($v_{rot}=4.6$~km\,s$^{-1}$) and the changes in the radial velocity during the exposure (2.3~km\,s$^{-1}$ for 300 s exposure and 3.1~km\,s$^{-1}$ for 400 s exposure)m using \texttt{fastRotBroad} routine from \texttt{PyAstronomy}. The models were injected into the in-transit spectra with the radial velocity shift which corresponds to the expected velocity of the planet in each of the orbital phases and 200 ~km\,s$^{-1}$ shift. The injected atmospheric models were calculated for the temperature of the planet ($T_{eq}$). In the final step, we cross-correlated the data with the binary mask of each of the studied atoms and molecules. 

Our final 3$\sigma$ upper limits, presented in Table \ref{tab:Upper_cc}, were calculated after cross-correlating the data with the binary mask of the studied atoms and molecules. For each of the species, we created the binary mask using previously calculated models with \texttt{PetitRADTRANS} for 1904 K by choosing the strongest lines, setting a strength minimum of 0.002\%. In contrast to the method presented by \citet{Casasayas_Barris_mascara_1_2022_ESPRESSO}, where only significant numbers of lines were used to calculate the upper limits, we used all the lines in the models. All of the chosen lines were given a value of 1, while the rest were set to 0. The number of lines used for each model can be found also in Table \ref{tab:Upper_cc}. The one-pixel dispersion was calculated as the standard deviation of the cross-correlation residuals for theoretical $K_p$, and velocities between $\pm$ 100~km\,s$^{-1}$, in the regions where we expect the atmospheric signal. We present two values, one calculated for the cross-correlation residuals with all the regions and the second after masking the regions where the Rossiter-McLaughlin effect has an impact. The values represent a minimum level of absorption which is needed to detect species using the cross-correlation method. 

For all the studied species, we were not able to recover the injected signal with the strength of the theoretical expected signal; thus, we did not calculate the line upper limit, as presented in \citet{Allart2017,2020_allart_wasp127_Na} and \citet{Casasayas_Barris_mascara_1_2022_ESPRESSO}. To carry out an additional recovery test, we injected the models with the strengths of 2x, 4x, and ten times the expected signals. In the case of H$_2$O and FeH, we were not able to recover any injected signal, which is probably related to the weak lines for those species at those temperatures in the visible part of the spectrum. For \ion{Ca}{I}, \ion{Cr}{i}, \ion{V}{i}, and \ion{Y}{i,} we were able to recover two times the signal with a low significance, as well as four times the signal with a high level of significance. For the rest of the species, we recovered the two times and four times signals with high significance. The S/N plots after injecting different signals can be found in Figs. \ref{fig:Injection_1} and \ref{fig:Injection_2}.

\begin{table}[h!]
\small\small
\centering
\caption{Upper limits of the cross-correlation detection, were calculated for the combination of the nights. A number of lines calculated for the strongest lines, with the strength of the lines: 0.002 \%.}
\begin{tabular}{l c c c }
\hline \hline
\\[-1em]
 Species  &Nr of lines& 3 $\sigma$ without / with mask [ppm]  \\ \hline
\\[-1em]
Ca I & 48  & 271/255\\
\\[-1em]
Cr I  &  82 & 375/336 \\
\\[-1em]
Fe I &  176 & 385/335\\
\\[-1em]
FeH  & 171  & 241/228 \\
\\[-1em]
H2O  & 332  & 275/269 \\
\\[-1em]
K I  & 42  & 268/249 \\
\\[-1em]
Na I & 22 & 311/253\\
\\[-1em]
Ti I  & 334  & 375/359\\
\\[-1em]
TiO-Exomol  & 1081  & 254/243\\
\\[-1em]
TiO-Plez  & 1082  & 293/274\\
\\[-1em]
V I  & 169  & 295/273\\
\\[-1em]
VO  & 1076  & 285/264\\
\\[-1em]
YI  & 41  & 278/265\\
\lasthline
\end{tabular}
\label{tab:Upper_cc}
\end{table}

%the injected models were broadened considering the resolution of the spectrograph using \textit{instrBroadGaussFast} routine from \textit{PyAstronomy} (czesla et al. 2019), considering the rotation of the tidally locked planet ($v_{rot}=4.6$~km\,s$^{-1}$), and the changes of the radial velocity during the exposure (2.3~km\,s$^{-1}$ for 300 s exposure and 3.1~km\,s$^{-1}$ for 400 s exposure) using \textit{fastRotBroad} routine from \textit{PyAstronomy} (czesla et al. 2019).

\section{Discussion and conclusions}\label{sec:discussion_conlucison}

%\begin{figure*}[h]
%\centering
%\includegraphics[width=0.9\textwidth]{FIG/V_VO_Y_cross.pdf}
%\caption{Cross-correlation results for KELT-14 for \ion{V}{i}, VO and \ion{Y}{i}. For each of the species we present the following plots. Left: Residual map, where red horizontal line represents beginning and end of the transit and the white tilted line represents the expected velocity of the planet. Center: $K_p$ map calculated for range of 0 to 300 km\,s$^{-1}$, the theoretical $K_p$ of the planets is represented by red horizontal line. Expected signal from the planet should be detected at $K_p=$185 km\,s$^{-1}$ and radial velocity 0 km\,s$^{-1}$. Right: SNR plot for theoretical $K_p$ value. The purple plot is raw SNR plot, and green plot is SNR after masking the regions where RM effect appear. }
%\label{fig:CC_1900_V_VO_Y}
%\end{figure*}

In this work, we study two nights of high-resolution observations of WASP-122b with the ESPRESSO spectrograph. By analyzing the Rossiter-McLaughlin effect, we were able to determine that WASP-122b is an aligned planet with an obliquity of $\lambda = 0.09^{+0.88}_{-0.90}$~deg. The results are consistent with the trend of planets orbiting stars with $T_{eff}$<6200 K which show a tendency to be aligned \citep{2010_winn}. More details are given in Fig. \ref{fig:Obliquity_demographics}. Our $v \sin{i}$ = 1.84 $\pm$ 0.08 km~s$^{-1}$ value is smaller than the two previously reported literature measurements of 3.3 $\pm$ 0.8 km\,s$^{-1}$ \citep{wasp-122_2016_Turner} and 7.7 $\pm$ 0.4 km\,s$^{-1}$ \citep{kelt-14_2016_Rodrigez}. This difference can be explained by the larger collective area, larger wavelength range, and higher resolution of the ESPRESSO data used in our analysis than that used by \citet{wasp-122_2016_Turner} (CYCLOPS data) and \citet{kelt-14_2016_Rodrigez} (CORALIE data).   
%\textcolor{red}{E: no I do not know. Maybe you just hjave better precision and the literature errors were underestimated. I would no tenter into why in the paper. You are missing a  reference for the second value.}

Additionally, to study the chemical composition of the atmosphere of WASP-122b, we used the transmission spectroscopy method. By looking at the transmission spectroscopic signal around single lines, we investigated the presence of the \ion{Na}{i} doublet, H$\alpha$, H$\beta$, and \ion{Li}{i}. The final results show no evidence of any of the studied atoms. 

Using the cross-correlation method, we searched for a list of atoms and the molecules in the atmosphere of WASP-122b: \ion{Ca}{i}, \ion{Cr}{i}, FeH, \ion{Fe}{i}, \ion{Fe}{ii}, H$_2$O, \ion{K}{i}, \ion{Mg}{i},  \ion{Na}{i},  \ion{Ti}{i}, TiO, \ion{V}{i}, VO, and \ion{Y}{i}. As in the case of single-line studies, we did not detect any signal with an S/N higher than 3$\sigma$.

The lack of detections can be explained by the absence or low abundances of all the studied species in the region of the atmosphere probed by our analysis. WASP-122b is a hot Jupiter; thus, we expect the existence of several molecules in its atmosphere, which show weak spectral lines in the visible part of the spectrum, but atomic species such as Na would also be expected. Alternatively, it is possible that the difficulties in the correction of the RM effect, which covers a broad range of the radial velocities where the planetary signal is expected, could explain these results. Additionally, the planet's atmosphere can be cloudy and most lines can be hidden or weakened under the cloud layer. Moreover, the low S/N of the observations makes it difficult to detect weaker signal coming from the atmosphere of this planet.

{The three planets, KELT-7b ($T_{eq}$ = 2048 $\pm$ 27 K, TSM $\sim$ 170, $M_p$ = 1.28 $\pm$ 0.18 M$_J$, $R_p$ = 1.533 $^{+0.046}_{-0.047}$ R$_J$, \citealt{KELT-7_Beryla}), WASP-19b ($T_{eq}$ = 2077 $\pm$ 34 K, TSM $\sim$ 170, $M_p$ = 1.139$^{+0.03}_{-0.02}$ M$_J$, $R_p$ = 1.410 $^{+0.017}_{-0.013}$ R$_J$, \citealt{wasp-19_Mancini}), and WASP-74b ($T_{eq}$ = 1910 $\pm$ 40, TSM $\sim$ 160, $M_p$ = 0.95 $\pm$ 0.06, $R_p$ = 1.56 $\pm$ 0.06 R$_J$, \citealt{wasp-74_West}) are  of a similar class to WASP-122b (TSM $\sim$ 180) in terms of all $T_{eq}$, radius, mass, and TSM. The atmosphere of KELT-7b was previously studied with the high-resolution spectrographs HORuS and HARPS-N showing no atmospheric signature (\citealt{2021_Stangret_6uhj}, \citealt{Tabernero_K7_2022}), while using the \textit{Hubble} Space Telescope (HST), \citealt{2022_Changeat} presented the detection of H$_2$O and H- in the atmosphere of this planet. The atmosphere of WASP-19b has been studied by several groups using both space- and ground-based spectrographs, showing the presence of TiO and H$_2$O (\citealt{w_19_2013}, \citealt{K_19_2016}, \citealt{sedaghati_2017}, \citealt{K_19_2021}, \citealt{2022_Changeat}).  In the case of WASP-74b, observations with HST revealed the presence of H$_2$O and CH$_4$ (\citealt{2022_Changeat}). Since all of those planets are in transition between hot and ultra-hot Jupiters, we can presume that for some of the atoms and molecules, it is extremely difficult to probe their atmospheric composition using high-resolution spectrographs since the abundance of those species can be in transition between natural and ionized states as well as undergoing dissociation. 

Future studies should focus on observations using the near-infrared part of the spectrum such as CRIRES$^+$, as well as space observations from HST or JWST. These observations could detect molecules such as water, CH$_4$, and a possible cloud layer, which could then be used for the atmospheric retrieval and help constrain the temperature of the upper atmosphere.

\begin{acknowledgements}

M.S. acknowledges the support of the PRIN INAF 2019 through the project ``HOT-ATMOS" and INAF GO Large Grant 2023 GAPS-2.

 JIGH acknowledge financial support from the Spanish Ministry of Science and Innovation (MICINN) project PID2020-117493GB-I00.;

We acknowledge funding from the European Research Council under the European Union’s Horizon 2020 research and innovation program under grant agreement No. 694513. This work was financed by Portuguese funds through FCT - Funda\c c\~ao para a Ci\^encia e a Tecnologia in the framework of the project 2022.04048.PTDC (Phi in the Sky). CJM also acknowledges FCT and POCH/FSE (EC) support through Investigador FCT Contract 2021.01214.CEECIND/CP1658/CT0001.;

This work was supported by Fundação para a Ciência e a Tecnologia (FCT) and Fundo Europeu de Desenvolvimento Regional (FEDER) via COMPETE2020 through the research grants UIDB/04434/2020, UIDP/04434/2020, 2022.06962.PTDC. O.D.S.D. is supported in the form of work contract (DL 57/2016/CP1364/CT0004) funded by FCT.;%Oliver

NCS was acknowledges the funding by the European Union (ERC, FIERCE, 101052347). Views and opinions expressed are however those of the author(s) only and do not necessarily reflect those of the European Union or the European Research Council. Neither the European Union nor the granting authority can be held responsible for them. This work was supported by FCT - Fundação para a Ciência e a Tecnologia through national funds and by FEDER through COMPETE2020 - Programa Operacional Competitividade e Internacionalização by these grants: UIDB/04434/2020; UIDP/04434/2020.; %Nuno Santos

This project has received funding from the Swiss National Science Foundation for project 200021\_200726. It has also been carried out within the framework of the National Centre of Competence in Research PlanetS supported by the Swiss National Science Foundation under grant 51NF40\_205606. The authors acknowledge the financial support of the SNSF.; %David Ehrenreich

R. A. is an SNSF Postdoctoral Fellow and acknowledges the SNSF support under the Post-Doc Mobility grant P500PT\_222212. This work was funded by the Institut Trottier de Recherche sur les Exoplan\`etes (iREx), and R.A. acknowledges support from the Trottier Family Foundation. This work has been carried out within the framework of the National Centre of Competence in Research PlanetS, supported by the Swiss National Science Foundation.;  %Allart

E.H.-C. acknowledges support from grant PRE2020-094770 under project PID2019-109522GB-C51 funded by the Spanish Ministry of Science and Innovation / State Agency of Research, MCIN/AEI/10.13039/501100011033, and by ‘ERDF, A way of making Europe’.; %Eva

This work made use of PyAstronomy \citep{Czesla_2019_pyastronomy} and of the VALD database, operated at Uppsala University, the Institute of Astronomy RAS in Moscow, and the University of Vienna. \texttt{ARoME} code is publicly available at \url{http://www.astro.up.pt/resources/arome/}. The Python translation of ARoME was made by A. Santerne.
\end{acknowledgements}

%%%%%%%%%%%%%% REFERENCES %%%%%%%%%%%%%%
\bibliographystyle{aa} % style aa.bst
\bibliography{biblio} % your references Yourfile.bib

\onecolumn
\begin{appendix} %First appendix

\section{Additional figures} \label{app:add_fig}

\begin{figure}[h]
\centering
\includegraphics[width=\textwidth]{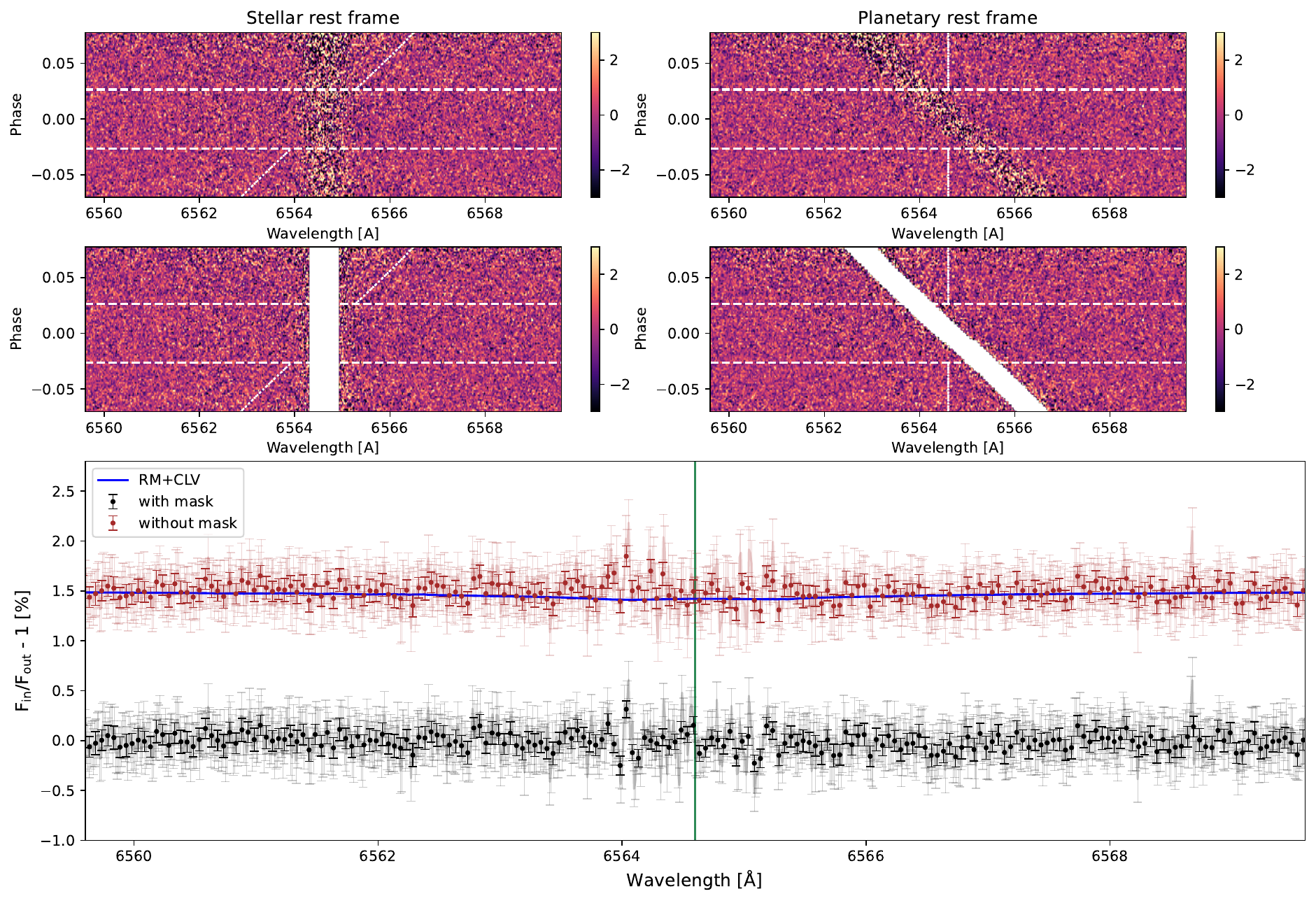}
\caption{Same as Fig. \ref{fig:transmission_Na}, but for H$\alpha$ (presented mask size $\pm$ 0.3 $\AA$).}
\label{fig:transmission_Ha}
\end{figure}

\begin{figure}[h]
\centering
\includegraphics[width=\textwidth]{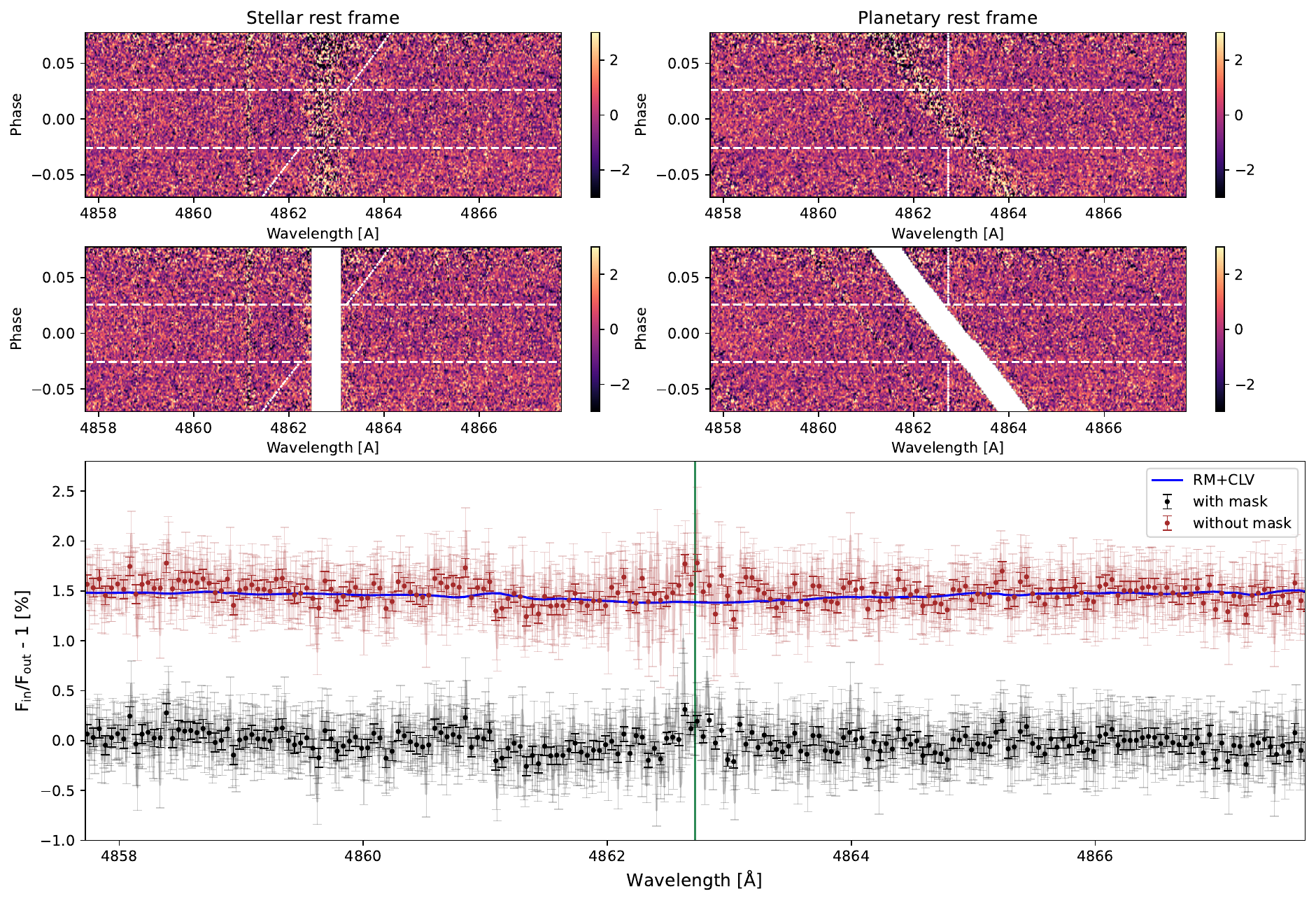}
\caption{Same as Fig. \ref{fig:transmission_Na}, but for H$\beta$ (presented mask size $\pm$ 0.3 $\AA$).}
\label{fig:transmission_Hb}
\end{figure}

\begin{figure}[h]
\centering
\includegraphics[width=\textwidth]{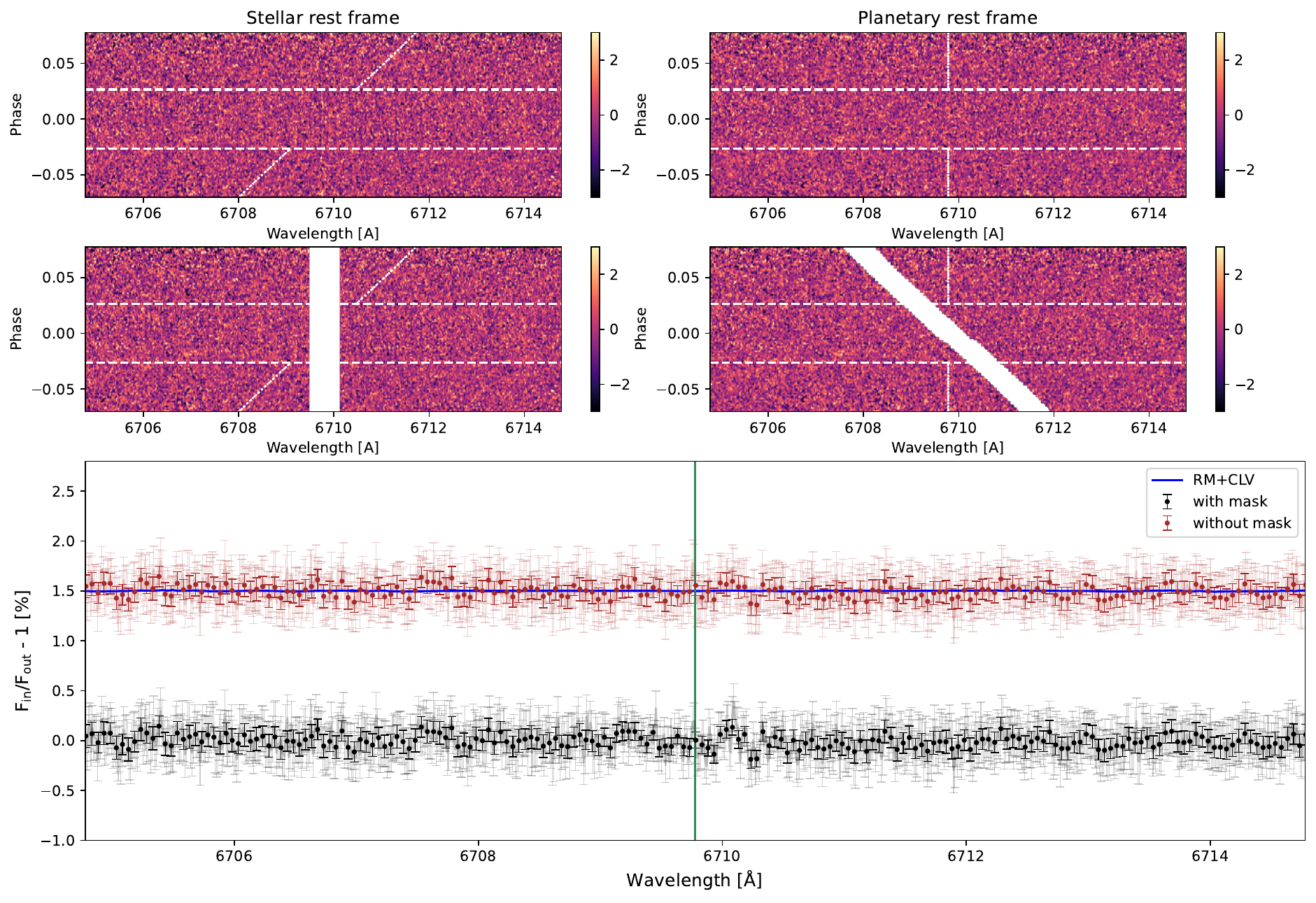}
\caption{Same as Fig. \ref{fig:transmission_Na}, but for \ion{Li}{i} (presented mask size $\pm$ 0.3 $\AA$).}
\label{fig:transmission_Li}
\end{figure}

\begin{figure}[h]
\centering
\includegraphics[width=\textwidth]{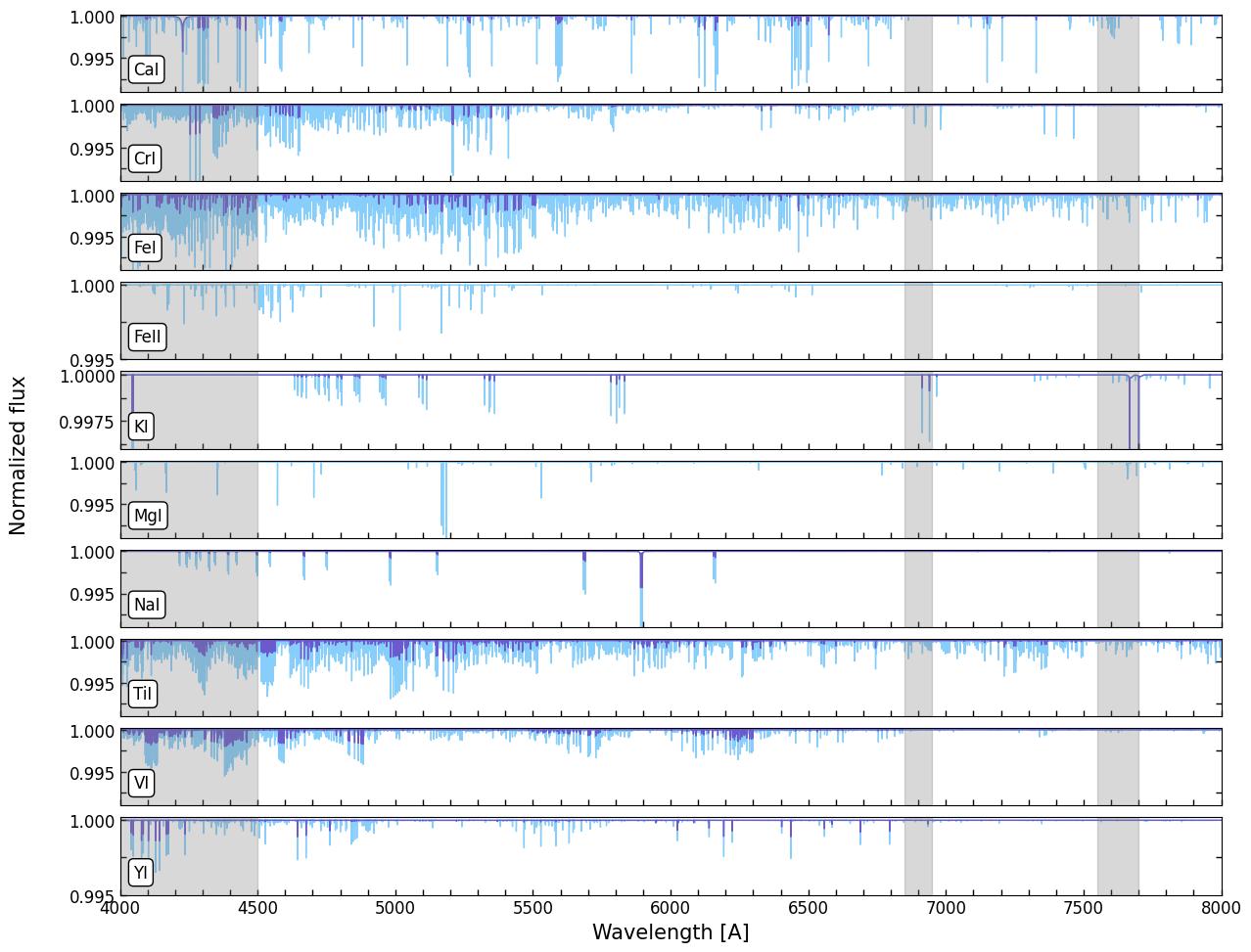}
\includegraphics[width=\textwidth]{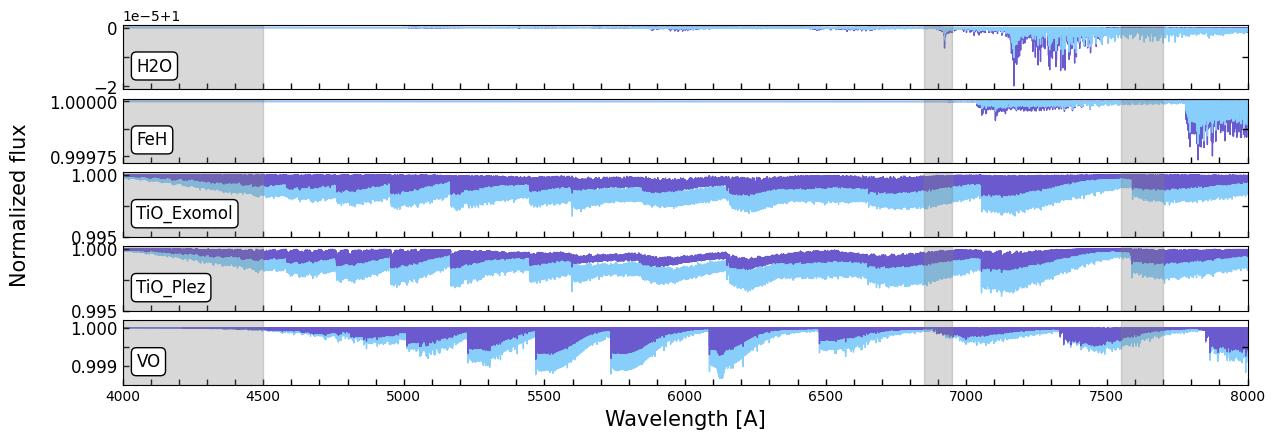}
\caption{Modeled spectrum of atoms (upper panel) and molecules (bottom panel):  \ion{Ca}{i} (Kurucz), \ion{Cr}{i} (Kurucz), FeH, \ion{Fe}{i} (Kurucz), \ion{Fe}{ii} (Kurucz), H2O (HITEMP), \ion{Mg}{i} (Kurucz), \ion{Na}{i} (Kurucz),  \ion{Ti}{i} (Kurucz), TiO (B. Plez, ExoMolOP), \ion{V}{i} (Kurucz), VO (ExoMolOP) ,and \ion{Y}{i} (Kurucz). Dark blue plots represent models calculated for 4000 K and light blue for 1900 K. In the case of \ion{Fe}{ii} and \ion{Mg}{i} there are no or few spectral lines for lower temperature, in this case, we only plot the models for higher temperature. In the grey vertical regions, we show the areas of the spectrum that were not used in the cross-correlation analysis. }
\label{fig:atoms_models}
\end{figure}

\begin{figure}[h]
\centering
\includegraphics[width=\textwidth]{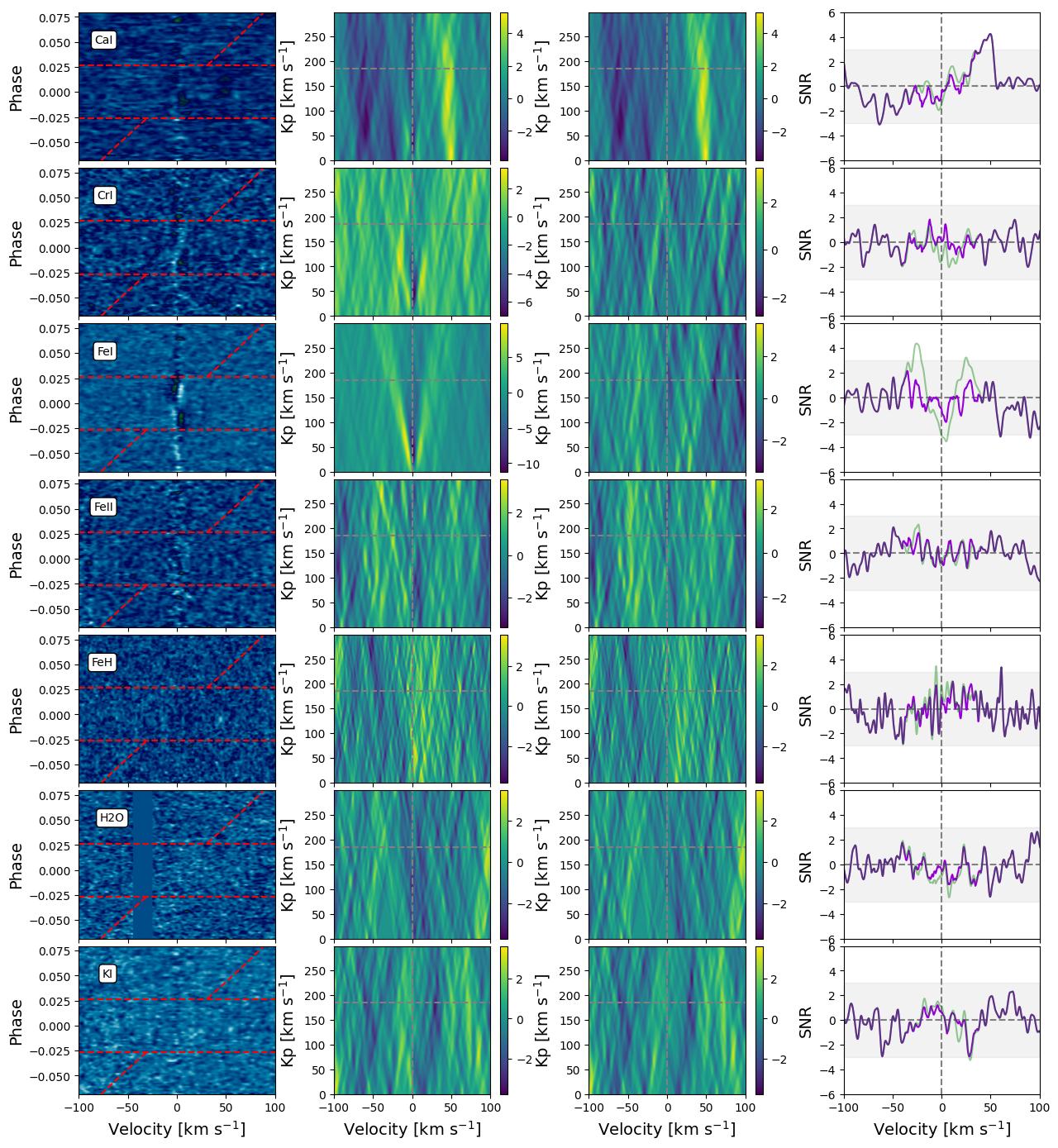}
\caption{Same as Fig. \ref{fig:CC_1900_1}, but for models of \ion{Ca}{i}, \ion{Cr}{i}, \ion{Fe}{i}, \ion{Fe}{ii}, FeH, H$_2$O, and \ion{K}{i} calculated for temperature of 4000 K.}
\label{fig:CC_4000_1}
\end{figure}

\begin{figure}[h]
\centering
\includegraphics[width=\textwidth]{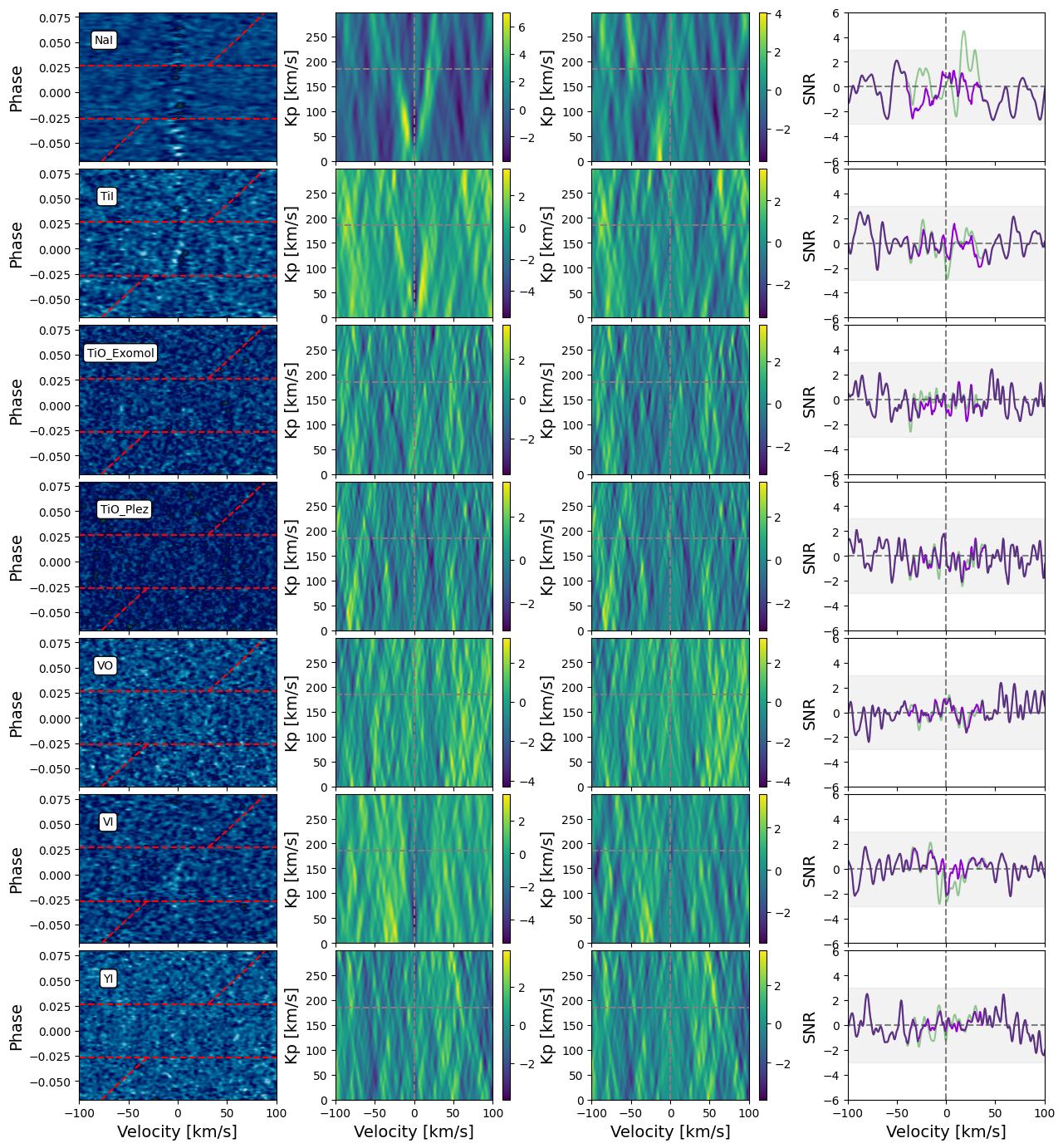}
\caption{Same as Fig. \ref{fig:CC_1900_1}, but for models of \ion{Na}{i}, \ion{Ti}{i}, TiO (Exomol), TiO (Plez), VO, \ion{V}{i}, and \ion{Y}{i} calculated for the temperature of 4000 K.}
\label{fig:CC_4000_2}
\end{figure}

\begin{figure}[]
\centering
\includegraphics[width=0.7\textwidth]{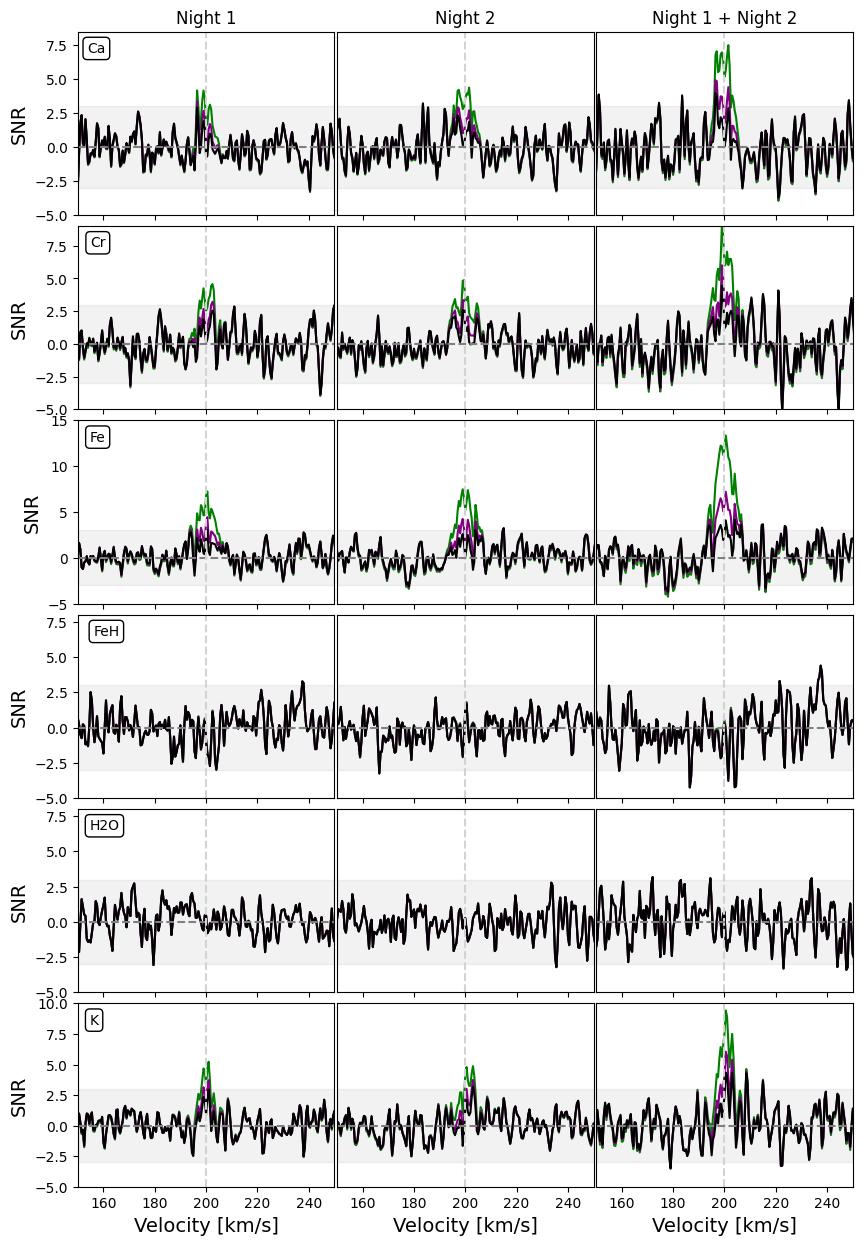}
\caption{S/N plots after injecting the signal of \ion{Ca}{i}, \ion{Cr}{i}, \ion{Fe}{i}, FeH, H$_2$O, and \ion{K}{i}. In the first column, we present the injection of the species into the data from Night 1, in the second column into Night 2, and in the third column, we present data after combining Night 1 and Night 2. The black plots represent the injected signal with an expected strength of the signal, the purple plot with a 2x of the expected signal, and the green with a 4x strength of the injected signal. The horizontal gray dashed line represents S/N = 0, and the vertical gray dashed line represents the radial velocity of 0 km/s. The gray regions represent S/N between -3 and 3.}
\label{fig:Injection_1}
\end{figure}

\begin{figure}[]
\centering
\includegraphics[width=0.7\textwidth]{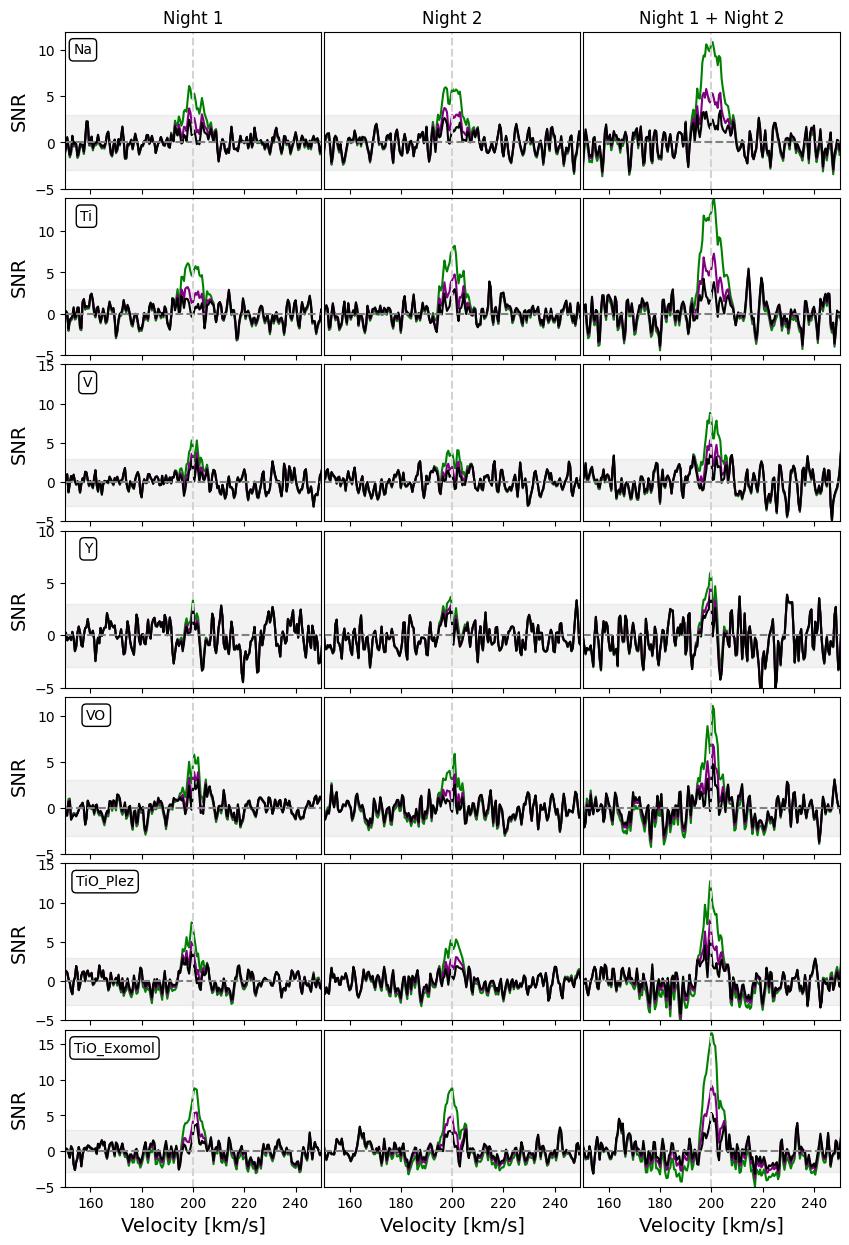}
\caption{Same as Fig. \ref{fig:Injection_1}, but for \ion{Na}{i}, \ion{Ti}{i}, \ion{V}{i}, \ion{Y}{i}, VO, and TiO. }
\label{fig:Injection_2}
\end{figure}

\begin{figure}[]
\centering
\includegraphics[width=0.9\textwidth]{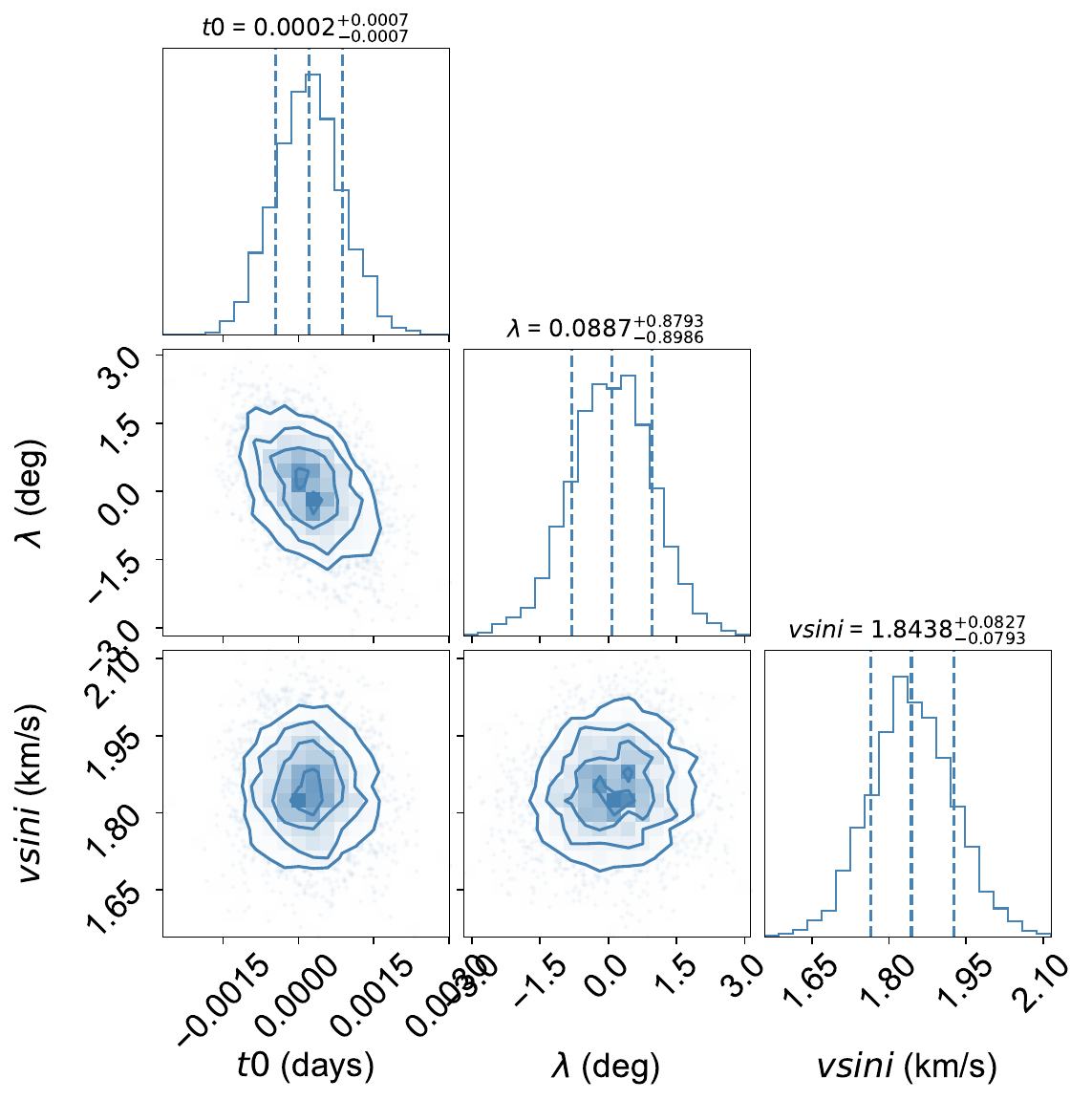}
\caption{Corner plot resulting from the \texttt{ARoME} + \texttt{emcee} fitting. The mid-transit time (t0), sky-projected spin-orbit angle ($\rm \lambda$), and the projected stellar rotational velocity ($\rm v\sin{i}$) parameters are left free during the fitting.}
\label{fig:RM Corners and chains}
\end{figure}

\end{appendix}

\end{document}